\documentclass[final,5p,times,twocolumn,12pt]{elsarticle}

\usepackage{url}
\usepackage{hyperref}
\hypersetup{
    colorlinks=true,
    linkcolor=blue,
    filecolor=magenta,      
    urlcolor=cyan,
}
\usepackage{xspace}
\usepackage{hyperref}
\usepackage{makecell}
\usepackage{multirow}
\usepackage{graphicx}
\usepackage{algorithm}
\usepackage{enumerate}
\usepackage{todonotes}
\usepackage{algpseudocode}
\usepackage{color, xcolor}
\usepackage{amsmath, amsfonts, amsthm}
\usepackage{balance}
\usepackage{pifont}
\usepackage{amsmath}
\usepackage{array}

\newcommand{\toolname}{OCP}
\usepackage{multirow}
\usepackage{bbm}
\usepackage{array}
\usepackage{CJKutf8}
\usepackage{graphicx} 
\usepackage{subfigure} 
\usepackage{fancybox}
\usepackage{algorithm}
\usepackage{algpseudocode}
\usepackage{pifont}
\newtheorem{definition}{\textbf{Definition}}

\usepackage{mathrsfs}
\usepackage{tikz}

\usepackage{threeparttable}
\usepackage{supertabular,booktabs}
\usepackage{rotating}
\usepackage{supertabular}

\usepackage{ulem} 

\newcommand{\finding}[2]{
\begin{center}
\fcolorbox{black}{gray!10}{\parbox{.97\linewidth}{
\textbf{Answer to RQ{#1}:}
{#2}
}}
\end{center}
}

\journal{The Journal of Systems and Software}

\begin{document}

\begin{frontmatter}

\title{Test Case Prioritization Using Partial Attention}

\author[]{Quanjun Zhang}
\ead{quanjun.zhang@smail.nju.edu.cn}
\author[]{Chunrong Fang\corref{cor1}}
\ead{fangchunrong@nju.edu.cn}
\author[]{Weisong Sun}
\ead{weisongsun@smail.nju.edu.cn}
\author[]{Shengcheng Yu}
\ead{yusc@smail.nju.edu.cn}
\author[]{Yutao Xu}
\ead{MF21320170@smail.nju.edu.cn}

\author[]{Yulei Liu}
\ead{1515999248@qq.com}

\cortext[cor1]{Corresponding author}
\address{State Key Laboratory for Novel Software Technology, Nanjing University, Nanjing, 210093, China}

\begin{abstract}

    Test case prioritization (TCP) aims to reorder the regression test suite with a goal of increasing the fault detection rate.
    Various TCP techniques have been proposed based on different  prioritization strategies. 
    Among them, the \textit{greedy-based} techniques are the most widely-used TCP techniques.
    However, existing \textit{greedy-based} techniques usually reorder all candidate test cases in prioritization iterations, resulting in both efficiency and effectiveness problems.
    In this paper, we propose a generic \textit{partial attention} mechanism, which adopts the previous priority values (i.e., the number of additionally-covered code units) to avoid considering all candidate test cases.
    Incorporating the mechanism with the \textit{additional-greedy} strategy, we implement a novel coverage-based TCP technique based on \textit{partition ordering} ({\toolname}).
    {\toolname} first groups the candidate test cases into different partitions and updates the partitions on the descending order.
    We conduct a comprehensive experiment on 19 versions of Java programs and 30 versions of C programs to compare the effectiveness and efficiency of {\toolname} with six state-of-the-art TCP techniques:
    \textit{total-greedy}, \textit{additional-greedy}, \textit{lexicographical-greedy}, \textit{unify-greedy}, \textit{art-based}, and \textit{search-based}.
    The experimental results show that {\toolname} achieves a better fault detection rate than the state-of-the-arts.
    Moreover, the time costs of {\toolname} are found to achieve 85\% -- 99\% improvement than most state-of-the-arts.


\end{abstract}

\end{frontmatter}

\section{Introduction}
\label{sec:intro}
During software maintenance and evolution, software engineers usually perform code modification due to the fixing of detected bugs, the adding of new functionalities, or the refactoring of system architecture \cite{2021Elsner, 2020Lam}.
Regression testing is conducted to ensure that the code modification does not introduce new bugs.
However, regression testing can be very time-consuming because of a large number of reused test cases \cite{wong1997study, 2015Gligoric2, 2018Zhang, 2019Cruciani}.
For example, Rothermel et al.~\cite{1999Rothermel} report that it takes seven weeks to run the entire test suite for an industrial project.
Besides, with the practices of rapid release \cite{2015Mantyla} and continuous integration \cite{2014Elbaum}, the available time for test execution recently keeps decreasing.
For example, Memon et al.~\cite{2017Memon} report that Google performs an amount of 800K builds and 150M test runs on more than 13K projects every day, consuming a lot of computing resources.

To address the overhead issues of regression testing,
test case prioritization (TCP) has become one of the most extensively investigated techniques \cite{sadri2022survey, do2020multi}.
Generally speaking, TCP reschedules the execution sequence of test cases in the entire test suite with the goal of detecting faults as early as possible.
Traditional TCP techniques \cite{wong1998effect, 2018khatibsyarbini, 2012yoo} usually involve an elementary topic, prioritization strategies, which incorporate test adequacy criteria (e.g., code coverage) to represent different behaviours of test cases.
In previous work, the most widely-investigated prioritization strategies are \textit{greedy-based} strategies \cite{1999Rothermel} (i.e., the \textit{total-greedy} and \textit{additional-greedy} strategies), which are generic for different coverage criteria.
Given a coverage criterion (e.g., statement or method coverage), the \textit{total-greedy} strategy selects the next test case yielding the highest coverage, whereas the \textit{additional-greedy} strategy selects the next test case covering the maximum code units not covered in previous iterations.
The recent empirical results show that although conceptually simple, the \textit{additional-greedy} technique has been widely recognized as one of the most effective TCP techniques on average in terms of fault detection rate \cite{2016Luo,2016Lu,2018Luo,2018Chi,2018Chen,2021Cheng}.


Compared with the \textit{total-greedy} strategy, the \textit{additional-greedy} strategy empirically performs outstandingly due to its feedback mechanism, where the next test case selection takes
into account the effect of already prioritized test cases \cite{2013Zhang, 2016Eghbali}.
However, there also exists a shortcoming in the \textit{additional-greedy} strategy.
Given a regression test suite $T$ with $n$ test cases, 
when selecting the $i$-th test case, the remaining $n - i + 1$ candidate test cases need to be updated.
Specifically, for each candidate test case, all not-yet-covered code units are examined, of which those covered by the candidate test case are identified.
The priority values of candidate test cases need to be measured 
based on the feedback binary states of each statement (i.e., covered or not covered).
As a result, the priority values of the candidate test cases in the previous iterations are lost and need to be recalculated in current iteration.


However, due to considering all candidate test cases in each iteration, the \textit{additional-greedy} strategy may suffer from the efficiency problem.
For example, consider 3 candidate test cases, expressed as $t_1$, $t_2$ and $t_3$, in a certain iteration, covering 4, 3 and 2 additional statements, respectively.
After the test case $t_1$ is selected, $t_2$ and $t_3$ need to be updated in the next iteration.
Ideally, the remaining test cases can cover a maximum of 3 additional statements in the next iteration, and only test case $t_2$ potentially satisfies the hypothesis.
If not, a further hypothesis that both test cases $t_2$ and $t_3$ have a maximum of 2 additionally-covered statements is considered, and so on.
As a result, the test cases that cover more statements in the previous iteration are more likely to maintain the advantage in the next iteration, as the test cases cannot cover more statements in the next iteration than in the previous iteration.
For example, updating the test cases covering no code units in previous iterations is unnecessary until the prioritization process repeats.
Thus, the \textit{additional-greedy} strategy, which considers all candidate test cases at once in each iteration, may bring redundant calculation in efficiency.

Besides, there is a high possibility of tie-occurring when considering all candidate test cases, may lead to a decrease performance in the effectiveness.
In the above example, a tie may occur if both $t_2$ and $t_3$ are considered at once (i.e., $t_2$ and $t_3$ has the highest coverage of statements not yet covered).
When facing a tie, the \textit{additional-greedy} strategy implicitly assumes that all remaining candidates are equally important, and selects one randomly.
However, previous empirical studies \cite{2016Eghbali} have shown that the probability of ties is relatively high in the \textit{additional-greedy} strategy, and the random tie-breaking strategy can be ineffective.
It can be observed that due to considering all candidate test cases in each iteration, the \textit{additional-greedy} strategy suffers from the both efficiency and effectiveness problems.

In this paper, to address the issues mentioned above, we propose a generic concept, partial attention mechanism, to avoid considering all candidate test cases with previous priority values  (i.e., the number of additionally-covered code units) .
We also apply the concept to the \textit{additional-greedy} strategy and implement a novel coverage-based TCP technique based on the notion of partition ordering ({\toolname}).
Our technique pays attention to the partial test cases instead of the whole candidate test set with the help of priority values calculated in the previous prioritization iteration.
The key idea of our technique is as follows: the priority values of the candidate test cases in the previous iteration can be regarded as a reference in the next iteration, so as to avoid considering all candidates at the same time.
To implement this idea, all candidate test cases are classified into different partitions based on their previous priority values.
Then among the candidates that have the highest priority value in the previous iteration, the one with the unchanged coverage of not-yet-covered code units is selected.
Likewise, if no test case meets the selection criterion, test cases with the second highest priority values are considered, and so on. 

We perform an empirical study to compare {\toolname} with six state-of-the-art TCP techniques in terms of testing effectiveness and efficiency on 19 versions of four Java programs, and 30 versions of five real-world Unix utility programs.
The empirical results demonstrate that {\toolname} can outperform state-of-the-arts in terms of fault detection rate.
{\toolname} is also observed to have much less prioritization time than most state-of-the-arts (except the \textit{total-greedy} strategy, a low bound control TCP technique) and the improvement can reach 85\% - 99\% on average.
We view our proposed technique as an initial framework to control the balance of full prioritization and partial prioritization during TCP, and believe more techniques can be derived based on our technique.
%


In particular, the contributions of this paper are as follows:
\begin{itemize}
  \item We propose the first notion of the partial attention mechanism that uses previous priority values to avoid considering all candidate test cases in TCP.
  \item We apply the partial attention mechanism to the \textit{additional-greedy} strategy, leading a novel coverage-based TCP technique based on partition ordering (\toolname).
  \item We conduct an empirical study to investigate the effectiveness and efficiency of the proposed technique compared to six state-of-the-art TCP techniques.
  \item We release the relevant materials (including source code, subject programs, test suites and mutants) used in the experiments for replication and future research \cite{myurl}.
\end{itemize}

The rest of this paper is organized as follows.
Section \ref{sec:bg&mv} reviews some background information and presents a motivation example.
Section \ref{sec:approach} introduces the proposed approach.
Section \ref{sec:exp} presents the research questions, and explains details of the empirical study.
Section \ref{sec:re&an} provides the detailed results of the study and answers the research questions.
Section \ref{sec:rw} discusses some related work, and Section \ref{sec:threats} discusses the threats to validity of our experiments.
Section \ref{sec:con} presents the conclusions and discusses future work.

\section{Background \& Motivation}
\label{sec:bg&mv}
In this section, we provide some background information about test case prioritization and a motivating example.

\subsection{Test Case Prioritization}
Test case prioritization (TCP) \cite{1999Rothermel} aims to reorder the test cases to maximize the value of an objective function (e.g., exposing faults earlier \cite{2000Elbaum} or reducing the execution time cost \cite{2009Zhang, 2012Mei}).
TCP problem is formally defined as follows:
\begin{definition}\textbf{Test Case Prioritization:} 
Given a test suite $T$, $PT$ is the set of its all possible permutations, and $f$ is an object function defined to map $PT$ to real numbers $\mathbb{R}$. 
The problem of TCP~\cite{1999Rothermel} is to find $P'\in PT$, such that $\forall P'',~ P''\in PT (P''\ne P'),~f(P')\ge f(P'')$.
\end{definition}

However, it is infeasible to obtain the fault detection capability of test cases before test execution. 
Therefore, some alternative metrics (e.g., structural coverage), which are in some way correlated with the fault detection rate, are adopted to guide the prioritization process instead \cite{1999Rothermel, 2017Wang}.
Among all metrics, code coverage is the most widely used one \cite{2016Luo,2019Lou}.
Intuitively, once a criterion is chosen, a specific prioritization strategy is used to order the test cases according to the chosen criterion, such as the \textit{greedy-based} strategies \cite{2000Elbaum}, \textit{search-based} strategies \cite{2007Li}, and \textit{art-based} strategies \cite{2009Jiang}.

\subsection{ A Motivating Example}

\begin{figure}[htbp]
\graphicspath{{graphs/}}
\centering
    \includegraphics[width=0.49\textwidth]{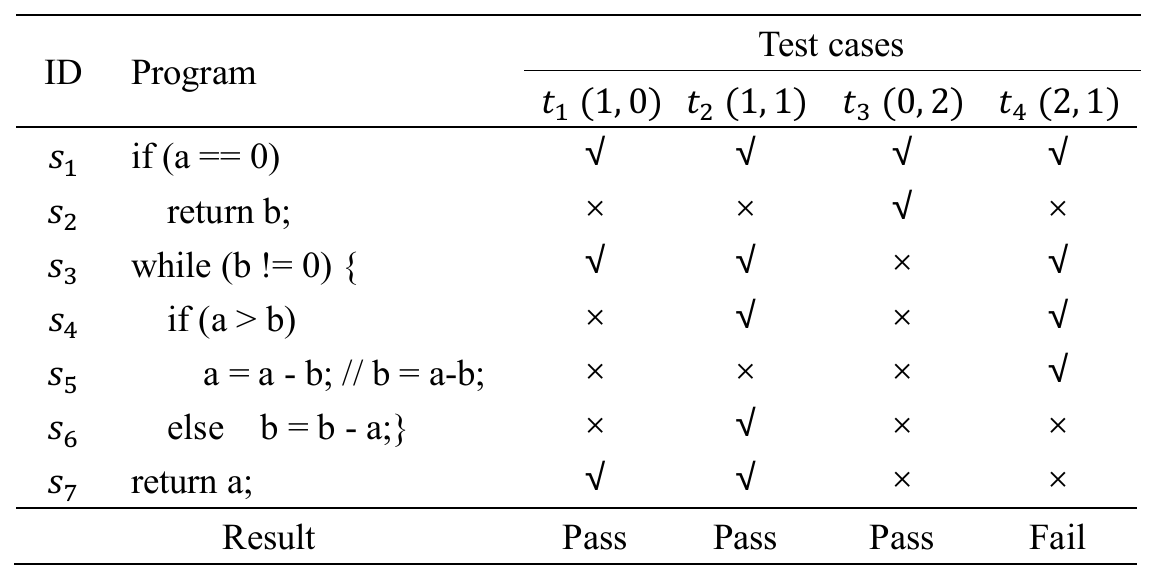}
    \caption{A motivating example}
    \label{FIG:example}
\end{figure}

To better illustrate the details of {\toolname}, Figure \ref{FIG:example} shows a piece of code with a fault in line $s_5$, which can be detected by the test case $t_4$. 
The code is a method that computes the greatest common divisor using the subtract-based version of Euclid’s algorithm \cite{2009Weimer, 2007Gazzola}. 
The source code is on the left and four test cases with their statement coverage information are on the right.

Before explaining the details of {\toolname}, we first review the steps that the \textit{additional-greedy} strategy takes to prioritize the four test cases.
In the first iteration, the \textit{additional-greedy} strategy chooses the test case $t_2$ with the maximum coverage.
To continue, in the second iteration, the \textit{additional-greedy} strategy selects the test case with the maximum coverage of not-yet-covered statements, e.g., $s_2$ and $s_5$.
The \textit{additional-greedy} strategy updates the coverage states for all remaining test cases and faces a tie where both $t_3$ and $t_4$ cover one of the not-yet-covered statements. 
In such a case, a random one (e.g., $t_3$ or $t_4$) will be selected.
In the third iteration, the \textit{additional-greedy} strategy searches for the test case, which yields the maximum coverage of statements that the first and the second test case have not covered, and $t_4$ or $t_3$ will be selected.
In other words, in each iteration, the \textit{additional-greedy} strategy selects the test case that provides the maximum coverage for the not-yet-covered statements. 
In the fourth iteration, the last test case $t_1$ is selected and this procedure continues until the ordering is complete.
As a result, the test sequence of the \textit{additional-greedy} strategy is $<t_2, t_3, t_4, t_1>$ or $<t_2, t_4, t_3, t_1>$ with the APFD values ranging from 0.375 to 0.625.
However, as discussed in Section \ref{sec:intro} the \textit{additional-greedy} strategy needs to consider all candidate test cases at each iteration, which may result in suboptimal performance in effectiveness and efficiency.
For example, in the second iteration, we need to update all the remaining test cases and also perform a random tie-breaking.



If {\toolname} is applied to the example, the first selected test case is $t_2$, which is the same as the \textit{additional-greedy} strategy.
However, in the second iteration, when facing a tie, {\toolname} prefers the fault-revealing test case $t_4$ as it covers more statements than $t_3$ in the first iteration.
In the third iteration, $t_3$ is updated first as it covers more statements than $t_1$ in the last iteration and found to cover one not-yet-covered statement $s_2$.
As no statement is covered by $t_1$ in the second iteration, we can observe $t_1$ cannot cover more statements in the next iteration.
Thus, $t_3$ is selected without updating $t1$, which leads to fewer calculations.
As a result, the test sequence of {\toolname} is $<t_2, t_4, t_3, t_1>$ with the APFD value reaching 0.625, which may result in a higher fault detection rate with lower prioritization time.
In recent years, the sizes of the regression test suites of modern industrial systems grow at a fast pace, and existing TCP techniques (e.g., the \textit{additional-greedy} strategy and its follow-ups) have become inadequate in efficiency \cite{2016Luo,2021Zhou}.
However, there exist little work to improve the \textit{additional-greedy} strategy efficiency while preserving the high effectiveness.

\section{Approach}
\label{sec:approach}
In this section, we introduce the details of test case prioritization by the partial attention mechanism.

\subsection{Partial Attention Mechanism}

Although the \textit{additional-greedy} strategy empirically performs outstandingly in terms of fault detection rate, there is a weakness in the feedback mechanism.
As discussed in Section \ref{sec:intro}, considering all candidate test cases in each prioritization iteration may result in redundant calculations and a high probability of tie-occurring.
In fact, only the binary states of the code units (i.e., covered or not covered) are fed back to the next iteration, and some valuable information(e.g., the previous priority values) is discarded.
In other words, the candidate test cases are independent of each other before each iteration, and the loss of previous priority values may lead to a decrease performance in the effectiveness and efficiency of TCP techniques.
As a result, the priority values of all candidate test cases need to be updated based on huge calculations.

Thus, to address the problem of considering all candidate test cases in each iteration, we attempt to adopt the feedback information from another perspective.
Specifically, the priority values in previous iterations are adopted to pay attention to partial candidate test cases.
The critical insight is that the number of additionally-covered code units is non-monotonically decreasing, as it cannot cover more code units in the next iteration.
Then, the priority values of candidate test cases can be stored in a well-designed structure (i.e., partition in Section \ref{pop}).
Meanwhile, the structure can be maintained in the next iteration, such that the more important test cases can be given more attention without additional calculations.
As a result, we propose a concept of the partial attention mechanism and apply the concept to state-of-the-art, the \textit{additional-greedy} strategy.

Suppose that test case $t_i$ and $t_j$ covers $s_{ik}$ and $s_{jk}$ not-yet-covered statements ($s_{ik} > s_{jk}$) at $k$-th iteration, respectively.
Thus, at the next iteration, $t_i$ is first updated and is found to cover $s_{i(k+1)}$ statements.
There may exist two possible situations: 
(1) $s_{i(k+1)} \leq s_{jk}$: $t_j$ needs to be updated and the number of covered statements is $s_{j(k+1)}$. 
If $s_{i(k+1)}$ equals $s_{j(k+1)}$, $t_i$ is preferred as it covers more statements than $t_j$ in the $i-$th iteration.
Otherwise, the test case covering more statements in $j$-th iteration is selected, which is identical to the \textit{additional-greedy} strategy.
(2) $s_{i(k+1)} > s_{jk}$: $t_i$ is selected without updating $t_j$.
Suppose the covered statements of selected test case at $k$-th iteration and $t_i$ are $S = \{s_1, s_2, ..., s_n\}$ and $S' = \{s'_1, s'_2, ..., s'_m\}$.
If $S \cup S' = \emptyset $, we have $s_{jk} = s_{jk}$, otherwise $s_{jk} > s_{jk}$.
Thus, we observe that as the selection steps iterate, the number of covered statements should be non-monotonically decreasing ($s_{jk} \geq s_{jl}$).
In such case, we can conform $s_{i(k+1)} > s_{j(k+1)}$ from $s_{i(k+1)} > s_{jk}$ and $s_{jk} \geq s_{jl}$ without updating $t_j$.

In conclusion, before covering all code units, if test case $t_i$ covers more code units than another test case $t_j$, $t_i$ is more likely to have a higher priority value.
Thus, instead of considering both $t_i$ and $t_j$, the more important one $t_i$ should be updated first.
If $t_i$ covers more code units than the theoretical best priority value of $t_j$ (i.e, the priority value at last iteration), $t_i$ will be selected.
Otherwise, the remaining $t_j$ is updated and compared with $t_i$.

\subsection{Partition Ordering based Prioritization}
\label{pop}

\begin{algorithm}[t]
\caption{Pseudocode of {\toolname}}
\scriptsize
\label{ALG:TCP}
\begin{algorithmic}[1]
    \renewcommand{\algorithmicrequire}{\textbf{Input:}}
    \renewcommand{\algorithmicensure}{\textbf{Output:}}

    \Require 
    $T$: $\{t_1,t_2,\cdots,t_n\}$ is a set of unordered test cases with size $n$;
    $U$: $\{u_1,u_2,\cdots,u_m\}$ is a set of code units with size $m$ in the program $P$
    \Ensure  
    $S$: a set of prioritized test cases
    
    \State $S  \leftarrow  \emptyset$
    \State $Candidates  \leftarrow  \emptyset$
    \State $priority \gets m$
    \Comment{the highest priority value}
    
    \For {each $j~(1 \leq j \leq m)$}
    	\State $UnitCover[j] \gets false$
    \EndFor
       
    \For {each $i~(1 \leq i \leq n)$}
    	\State $Candidates \gets Candidates \succ \langle t_i, priority \rangle$
    \EndFor
    
    \While {$|Candidates| > 0$}
    	\State $maximum \gets -1$
    	\For {each $\langle t_i, temp\_priority \rangle ~ \epsilon ~ Candidates$}
    		\If{$temp\_priority  == priority$}
    			\State $temp\_num \gets 0$
    			\For {each $j~(1 \leq j \leq m)$}
    				\If{$Cover[i, j]$ and not $UnitCover[j]$}
    					\State $temp\_num \gets temp\_num + 1$
    				\EndIf
    			\EndFor
    		\EndIf
    			\If{$temp\_num = priority$}
    				\State $temp\_Candidates \succ \langle t_i, priority \rangle$
    			\EndIf
        \EndFor

        \If{$priority > 0$} 
            \If{$|temp\_Candidates| > 0 $}
            	\State $\langle t_k, priority \rangle \gets TieSelect(temp\_Candidates)$
            	\State $S \gets S \succ \langle t_k \rangle$
            	\State $Candidates \gets S \setminus \langle t_k, priori \rangle$
            	
            	\For {each $j~(1 \leq j \leq m)$}
            		\If{$Cover[i, j]$ and not $UnitCover[j]$}
            			\State $UnitCover[j] \gets false$
    				\EndIf
				\EndFor
            \Else
            	\State $priority \gets priority - 1$
            \EndIf
        \Else
			\For {each $j~(1 \leq  j \leq m)$}
				\State $UnitCover[j] \gets false$
			\EndFor
        \EndIf

    \EndWhile
    \State \Return $S$

\end{algorithmic}
\end{algorithm}

In our work, we view the partial attention mechanism as a general concept that can be applied to different prioritization strategies using different coverage criteria. 
For example, in the \textit{lexicographical-greedy} strategy, the test cases covering the code units fewer covered in the previous iteration should be preferred, as they have a higher probability to cover these code units in the next iteration.
As \textit{greedy-based} strategies are the most widely-adopted prioritization strategies \cite{1999Rothermel, 2013Zhang}, and the \textit{additional-greedy} strategy is considered to be one of the most effective TCP techniques in terms of fault detection capacity \cite{2016Luo,2016Lu, 2021Cheng,2019Luo, li2021aga}.
We apply the partial attention mechanism to the \textit{additional-greedy} strategy and implement a simple greedy strategy to instantiate the function based on partition ordering.
Generally speaking, all candidate test cases are grouped into different partitions based on their previous priority values and higher partitions are updated preferentially.

Given a program under test $U = \{u_1, u_2, \cdots, u_m\}$ containing $m$ code units, and a test suite $T = \{t_1, t_2, \cdots, t_n\}$ containing $n$ test cases, Algorithm \ref{ALG:TCP} describes the pseudocode of our proposed method.
At each iteration, all the candidate test cases are sorted based on the priority values (i.e., the number of additionally-covered code units) and the ones with the same priority value are adjacent to each other. 
We then group the candidate test cases according to their priority values into $p$ partitions, indexed from left to right by $1, 2, ..., p$, such that all the candidate test cases within a partition have the same priority value.
Based on these partitions, we form the vector $v = [v_1, v_2, ..., v_p]$ where $v_i$ indicates priority value in the $i-th$ partition.
In the next iteration, we then update the candidate test cases in partition $p$ and group them into new partitions according to their updated priority values (i.e., the number of additionally-covered code units in the next iteration).
If there exists a test case from the partition $p$ that falls into a new partition $j~ (v_j > v_{p-1})$, the test case is selected.
Because the test cases from the partition $i~(i \leq p-1)$ can not be updated into a partition with a higher index $j~(v_j > v_{p-1})$.
Otherwise, we update the partition from right to left according to the value $v_i$ until a test case is selected.
In the worst case, we may update all the partitions if no test case is selected, which is identical to the \textit{additional-greedy} strategy.

Specifically, we use a Boolean array $Cover[i, j] ~ (1\leq i \leq n, ~1\leq j \leq m)$ to identify whether the test case $t_i$ covers the code unit $u_j$ or not.
We use another Boolean array $UnitCover[j] ~ (1 \leq j \leq m)$ to denote whether the code unit $u_j$ has been covered by the already selected test cases or not.
We set the value of $UnitCover[j] ~ (1 \leq j \leq m)$ to be $false$.
Similarly, we use a variable $priority$ to denote the largest priority value for all candidate test cases.
Meanwhile, we set the value of $priority$ to be $m$.
Besides, we use a set $Candidates$ to denote all remaining test cases and their corresponding priority values.
Initially, we add the whole test suite to $Candidates$ with the default priority value $m$ (i.e., the number of code units).

In Algorithm \ref{ALG:TCP}, lines 1-9 perform initialization, and lines 10--43 prioritize the test cases.
In the main loop from line 10 to line 43, each iteration attempts to find a test case with the given priority value and add it to the prioritized test set $S$.
In particular, lines 12-24 calculate the sum of units covered by the test cases with the highest previous priority value and add the ones that maintain the advantage into the candidate test set $temp\_Candidates$.
Before choosing the next test case, our approach examines whether or not there are any code units that are not covered by the test cases in $S$.
If all code units have been covered, the remaining candidate test cases are prioritized by restarting the previous process (lines 39-41).
Otherwise, we select the test case in $temp\_Candidate$ with highest previous priority values as the next one and update the cover status of all code units (lines 26-37).
If no test case is selected, we further update the second partition, and so on (line 36).
This process is repeated until all test cases in $Candidates$ have been added to $S$.
It is worth noting that although a partial attention mechanism is adopted in our approach, there is also a small possibility that a tie occurs, e.g., more than one test case in $temp\_Candidate$ with highest previous priority values.
In such a case, similar to the \textit{additional-greedy} strategy, our approach performs a random tie-breaking.

\section{Experiment}
\label{sec:exp}

In this section, we present our empirical study in detail, including the research questions, some variables, subject programs and the experimental setup.

\subsection{Research Questions}
The empirical study is conducted to answer the following research questions.

\begin{description}
    \item[RQ1] How does the effectiveness of {\toolname} compare with state-of-the-art techniques, in terms of fault detection rate?
    \item[RQ2] How does the granularity of code coverage impact the comparative effectiveness of {\toolname}?
    \item[RQ3] How does the granularity of test cases impact the comparative effectiveness of {\toolname}?
    \item[RQ4] How does the efficiency of {\toolname} compare with state-of-the-art techniques, in terms of execution time?
\end{description}

\subsection{Independent Variables}
\subsubsection{Prioritization techniques}

Although the proposed generic strategies can work with any coverage criteria, we implement {\toolname} based on basic structural coverage criteria due to their popularity \cite{2016Lu, 2013Zhang,2019Lou, 2009Jiang,2019Luo}.
We select the six state-of-the-art coverage-based TCP techniques that have been widely used in previous TCP studies \cite{2016Luo,2016Lu,2019Luo}: \textit{total-greedy} \cite{1999Rothermel}, \textit{additional-greedy} \cite{1999Rothermel}, \textit{unified-greedy} \cite{2013Zhang,2014Hao}, \textit{lexicographical-greedy} \cite{2016Eghbali}, \textit{art-based} \cite{2009Jiang}, and \textit{search-based} \cite{2007Li}.

The \textit{total-greedy} technique prioritizes test cases based on the descending number of code units covered by those test cases.
The \textit{additional-greedy} technique chooses each test case from the candidate test set such that it covers the largest number of code units not yet covered by the previously selected test cases.
Similarly, the \textit{unified-greedy} technique selects the test case with the highest sum of the probabilities that units covered by the test case contain undetected faults, while the \textit{lexicographical-greedy} technique selects the test case with the maximum coverage of one-time-covered code units.
Likewise, if a tie occurs, code units that are covered twice are considered, and so on.
The \textit{art-based} technique selects each test case from a random candidate test set such that it has the greatest maximum distance from the already selected test cases.
Finally, the \textit{search-based} technique considers all permutations as candidate solutions, and uses a meta-heuristic search algorithm to guide the search for a better test execution order \cite{2007Li}.
Depending on prioritization strategies, these TCP techniques are grouped into three categories and the details are presented in Table \ref{TAB:TCPtechniques}.

For the \textit{total-greedy}, \textit{additional-greedy}, \textit{art-based} and \textit{search-based} techniques, we directly use the source code released by existing work \cite{2018Chen,2020Huang}.
Meanwhile, the implementation of the \textit{unified-greedy} technique is not publicly  available and the \textit{lexicographical-greedy} technique is implemented in other language (i.e., Matlab).
Thus, we implement the \textit{unified-greedy} and \textit{lexicographical-greedy} techniques according to their paper carefully.
For the \textit{unified-greedy} technique, we select the basic model (i.e., Algorithm 1 in \cite{2013Zhang}) in our work, as the extended model requires multiple coverage of code units by given test cases, which is beyond the scope of our work.
We also select the default configuration (i.e., Algorithm 2 in \cite{2016Eghbali}) for the \textit{lexicographical-greedy} technique, as it achieve a great balance between fault detection rate and prioritization time.

 

\begin{table}[htbp]
\centering
\scriptsize
\caption{Studied TCP techniques}
 \label{TAB:TCPtechniques}
    \setlength{\tabcolsep}{0.5mm}{
    \begin{tabular}{llll}
    \hline
        \textbf{Mnemonic}  &\textbf{Description}  & \textbf{Category} &\textbf{Reference} \\
        \hline
        $\textit{TCP}_\textit{tot}$ & \textit{total-greedy} test  prioritization & greedy-based &\cite{1999Rothermel}  \\
        $\textit{TCP}_\textit{add}$ & \textit{additional-greedy} test  prioritization& greedy-based &\cite{1999Rothermel}\\
        $\textit{TCP}_\textit{unif}$ & \textit{unified-greedy} test  prioritization & greedy-based&\cite{2013Zhang,2014Hao}\\
        $\textit{TCP}_\textit{lexi}$ & \textit{lexicographical-greedy} test  prioritization & greedy-based&\cite{2016Eghbali}\\
        $\textit{TCP}_\textit{art}$ & \textit{art-based} test  prioritization& similarity-based &\cite{2009Jiang}\\
        $\textit{TCP}_\textit{search}$ & \textit{search-based} test  prioritization& search-based &\cite{2007Li}\\
 
        $\textit{TCP}_\textit{\MakeLowercase{\toolname}}$ & our proposed technique {\toolname}& greedy-based &This study\\
        \hline
 \end{tabular}}
\end{table}

 \subsubsection{Code Coverage Granularity}
In traditional TCP studies \cite{2016Luo,2013Zhang}, the coverage granularity is generally considered to be a constituent part of the prioritization techniques.
To enable sufficient evaluations, we attempt to investigate generic prioritization strategies with various structural coverage criteria (i.e., the statement, branch, and method coverage granularities).

 \subsubsection{Test Case Granularity}
For the subject programs written in Java, we consider the test case granularity as an additional factor in the prioritization techniques.
Test case granularity is at either the test-class or the test-method granularity.
Specifically, given a Java program, a JUnit test class file refers to a test case at test-class granularity, while each test method in the file refers to a test case at test-method granularity.
In other words, a test case at the test-class granularity generally involves a number of test cases at the test-method granularity.
For C subject programs, the actual program inputs are the test cases.

\subsection{Dependent Variables}
To evaluate the effectiveness of different TCP techniques, we adopt the widely-used APFD (\textit{average percentage faults detected}) as the evaluation metric for fault detection rate ~\cite{1999Rothermel}.
Given a test suite $T$, with $n$ test cases, $P'$ is a permutation of $T$.
Then the APFD value for $P'$ is defined by the following formula:
\begin{equation}
\footnotesize
	APFD=1-\frac{\sum_{i=1}^{m}{TF_i}}{n*m}+\frac{1}{2n}
\end{equation}
where, $m$ denotes the total number of detected faults and $TF_i$ denotes the position of first test case that reveals the fault $i$.

\subsection{Subject Programs, Test Suites and Faults}
To enable sufficient evaluations, we conduct our study on 19 versions of four Java programs (i.e., eight versions of \textit{ant}, five versions of \textit{jmeter}, three versions of \textit{jtopas}, and three versions of \textit{xmlsec}), which are obtained from the \textit{Software-artifact Infrastructure Repository} (SIR)~\cite{2005Do,sir}.
Meanwhile, 30 versions of five real-life Unix utility programs written in C language (six versions of \textit{flex}, \textit{grep}, \textit{gzip}, \textit{make} and \textit{sed}) are also adopted, which are downloaded from the \textit{GNU FTP server}~\cite{gnu}.
Both the Java and C programs have been widely utilized as benchmarks to evaluate TCP tchniques~\cite{2013Zhang,2016Eghbali,2009Jiang,2016Henard}.
Table \ref{TAB:programs} lists all the subject programs and the detailed statistical information.
In Table \ref{TAB:programs}, for each program, columns 3 to 6 summarize the version, size, number of branches,  number of methods, respectively.

Each version of the Java programs has a JUnit test suite that is developed during the program's development.
These test suites have two levels of test-case granularity: the test-class and the test-method.
The numbers of JUnit test cases are shown in the \textbf{\#Test} column:
The data is presented as $x~(y)$, where $x$ is the number of test cases at test-method granularity, and $y$ is the number of test cases at test-class granularity.
The test suites for the C programs are collected from the SIR~\cite{2005Do,sir}.
The number of tests cases in each suite is also shown in the \textbf{\#Test} column of Table \ref{TAB:programs}.

The faults contained in each version of the programs are produced based on mutation analysis \cite{2019Papadakis, 2019Zhang}.
Although some seeded faults of programs are available from SIR, previous research has confirmed that the seeded ones are easily detected and small in size.
Meanwhile, mutation faults have previously been identified as suitable for simulating real program faults~\cite{2005Andrews,belli2006basic,2005Do_a,2014Just,belli2016model} and have been widely applied to various TCP  evaluations~\cite{1999Rothermel,2016Luo,2016Lu,2013Zhang,2000Elbaum,2019Luo,2016Henard}.
Thus, for both C and Java programs, mutation faults are introduced to evaluate the performance of the different techniques.
The details of these operators are presented in Table \ref{TAB:Mutators}.
For C programs, we obtain the mutants from previous TCP studies \cite{2016Henard, 2006Andrews}, which are produced using seven mutation operators.
For Java programs, we use eleven mutation operators from the ``NEW\_DEFAULTS'' group of the PIT mutation tool~\cite{pit} to generate mutants.
Specifically, we generate mutants (i.e., faulty versions) by seeding all mutation operators into the subject programs automatically.
Then we run the available test suite against each mutant.
The mutant is killed if there exist any test that produces inconsistent test outcomes between the original and faulty version,
otherwise the mutant is lived.
We select all killed mutants to evaluate the fault detection rate of TCP techniques.

Meanwhile, according to existing studies \cite{2020Huang,2016Henard},  the subsuming mutants identification (SMI) technique \cite{Papadakis2016} is adopted to remove the duplicate and subsuming mutants from all killed mutants.
The number of subsuming mutants used in our experiment is presented in the \textbf{\#Subsuming\_Mutant} column.
It's worth noting that the subsuming faults are classified as test-class level and test-method level for the Java programs.

\begin{table}[t]
\centering
\scriptsize
\caption{Statistics on Mutation Operators}
 \label{TAB:Mutators}
    \setlength{\tabcolsep}{3.5mm}{
    \begin{tabular}{lll}
    \hline
        \textbf{Language}  &\textbf{Operators} &\textbf{Descriptions} \\
        \hline
        \multirow{11}*{Java} & CB & \textit{conditionals boundary} \\
                            & IC & \textit{increments} \\
                            & IN & \textit{invert negatives} \\
                            & MA & \textit{math} \\
                            & NC & \textit{negate conditionals} \\
                            & VM & \textit{void method calls} \\
                            & ER & \textit{empty returns} \\
                            & FR & \textit{false returns} \\
                            & TR & \textit{true returns} \\
                            & NR & \textit{null returns} \\
                            & PR & \textit{primitive returns} \\

        \hline
        \multirow{7}*{C} & SD & \textit{statement deletion} \\
                        & UI & \textit{unary insertion} \\
                        & CR & \textit{constant replacement} \\
                        & AR & \textit{arithmetic operator replacement} \\
                        & LR & \textit{logical operator replacement} \\
                        & BR & \textit{bitwise logical operator replacement} \\
                        & RR & \textit{relational operator replacement} \\
        
        \hline
 \end{tabular}}
\end{table}

\begin{table*}[!t]
\scriptsize
\centering
\caption{Subject program details}
 \label{TAB:programs}
 \setlength{\tabcolsep}{1.4mm}{
    \begin{tabular}{c|c|r|r|r|r|r|r|r|r|r|c|c}
    \hline
        \multirow{2}*{\textbf{Language}} &\multirow{2}*{\textbf{Program}} &\multirow{2}*{\textbf{Version}} &\multirow{2}*{\textbf{KLoC}} &\multirow{2}*{\textbf{\#Branch}} &\multirow{2}*{\textbf{\#Method}} &\multirow{2}*{\textbf{\#Class}} &\multicolumn{2}{c|}{\textbf{\#Test\_Case}} &\multicolumn{2}{c|}{\textbf{\#Mutant}} &\multicolumn{2}{c}{\textbf{\#Subsuming\_Mutant}} \\\cline{8-13}
        &&&&&&&\textbf{\#T\_Class} &\textbf{\#T\_Method} &\textbf{\#All} &\textbf{\#Detected} &\textbf{\#SM\_Class} &\textbf{\#SM\_Method}\\
    \hline

\multirow{14}*{Java}	
&$ant\_v1$	&v1\_9	&25.80 	&5,240	&2,511	&228	&34 (34)	&137 (135)	&6,498	&1,332	&59	&32	\\
&$ant\_v2$	&1.4	&39.70 	&8,797	&3,836	&342	&52 (52)	&219 (214)	&11,027	&2,677	&90	&47	\\
&$ant\_v3$	&1.4.1	&39.80 	&8,831	&3,845	&342	&52 (52)	&219 (213)	&11,142	&2,661	&92	&47	\\
&$ant\_v4$  &1.5        &61.90  &11,743   &5,684  &532    &102 (100)    &521 (503)  &14,834 &6,585  &192 &88 \\
&$ant\_v5$  &1.5.2      &63.50  &141,76   &5,802  &536    &105 (103)    &557 (543)  &17,826 &6,230  &211 &91 \\
&$ant\_v6$  &1.5.3      &63.60  &141,68   &5,808  &536    &105 (102)    &559 (537)  &17,808 &6,255  &92 &91 \\
&$ant\_v7$  &1.6 beta    &80.40  &17,164  &7,520  &649    &149 (149)    &877 (866)  &22,171 &9,094  &284 &119 \\
&$ant\_v8$  &1.6 beta2   &80.40  &17,746  &7,524  &650    &149 (149)    &879 (867)  &22,138 &9,068  &226 &119 \\

\cline{2-13}												
&$jmeter\_v1$	&v1\_7\_3	&33.70 	&3,815	&2,919	&334	&26 (21)	&78 (61)	&8,850	&573	&38	&20	\\
&$jmeter\_v2$	&v1\_8   &33.10 	&3,799	&2,838	&319	&29 (24)	&80 (74)	&8,777	&867	&37	&22	\\
&$jmeter\_v3$	&v1\_8\_1 &37.30 	&4,351	&3,445	&373	&33 (27)	&78 (77)	&9,730	&1,667	&47	&25	\\
&$jmeter\_v4$	&v1\_9\_RC1  &38.40 	&4,484	&3,536	&380	&33 (27)	&78 (77)	&10,187	&1,703	&47	&25		\\
&$jmeter\_v5$	&v1\_9\_RC2  &41.10 	&4,888	&3,613	&389	&37 (30)	&97 (83)	&10,459	&1,651	&53	&29		\\
\cline{2-13}												
&$jtopas\_v1$	&0.4	&1.89 	&519	&284	&19	&10 (10)	&126 (126)	&704	&399	&29	&9	\\
&$jtopas\_v2$	&0.5.1	&2.03 	&583	&302	&21	&11 (11)	&128 (128)	&774	&446	&34	&10	\\
&$jtopas\_v3$	&0.6	&5.36 	&1,491	&748	&50	&18 (16)	&209 (207)	&1,906	&1,024	&57	&16	\\
\cline{2-13}												
&$xmlsec\_v1$	&v1\_0\_4    &18.30 	&3,534	&1,627	&179	&15 (15)	&92 (91)	&5,501	&1,198	&32	&12	\\	
&$xmlsec\_v2$	&v1\_0\_5D2  &19.00 	&3,789	&1,629	&180	&15 (15)	&94 (94)	&5,725	&1,204	&33	&12	\\	
&$xmlsec\_v3$	&v1\_0\_71   &16.90 	&3,156	&1,398	&145	&13 (13)	&84 (84)	&3,833	&1,070	&27	&10	\\	
	 \hline												
\multirow{30}*{C}	&$flex\_v0$	&2.4.3	&8.96	&2,005	&138	&--	&\multicolumn{2}{c|}{500}		&--	&--	&\multicolumn{2}{c}{--}		 \\
	&$flex\_v1$	&2.4.7	&9.47	&2,011	&147	&--	&\multicolumn{2}{c|}{500}		&13,873	&6,177	&\multicolumn{2}{c}{32}		 \\
	&$flex\_v2$	&2.5.1	&12.23	&2,656	&162	&--	&\multicolumn{2}{c|}{500}		&14,822	&6,396	&\multicolumn{2}{c}{32}		 \\
	&$flex\_v3$	&2.5.2	&12.25	&2,666	&162	&--	&\multicolumn{2}{c|}{500}		&775	&420	&\multicolumn{2}{c}{20}		 \\
	&$flex\_v4$	&2.5.3	&12.38	&2,678	&162	&--	&\multicolumn{2}{c|}{500}		&14,906	&6,417	&\multicolumn{2}{c}{33}		 \\
	&$flex\_v5$	&2.5.4	&12.37	&2,680	&162	&--	&\multicolumn{2}{c|}{500}		&14,922	&6,418	&\multicolumn{2}{c}{32}		 \\
	\cline{2-13}												
	&$grep\_v0$	&2.0	&8.16	&3,420	&119	&--	&\multicolumn{2}{c|}{144}		&--	&--	&\multicolumn{2}{c}{--}		 \\
	&$grep\_v1$	&2.2	&11.99	&3,511	&104	&--	&\multicolumn{2}{c|}{144}		&23,896	&3,229	&\multicolumn{2}{c}{56}		 \\
	&$grep\_v2$	&2.3	&12.72	&3,631	&109	&--	&\multicolumn{2}{c|}{144}		&24,518	&3,319	&\multicolumn{2}{c}{58}		 \\
	&$grep\_v3$	&2.4	&12.83	&3,709	&113	&--	&\multicolumn{2}{c|}{144}		&17,656	&3,156	&\multicolumn{2}{c}{54}		 \\
	&$grep\_v4$	&2.5	&20.84	&2,531	&102	&--	&\multicolumn{2}{c|}{144}		&17,738	&3,445	&\multicolumn{2}{c}{58}		 \\
	&$grep\_v5$	&2.7	&58.34	&2,980	&109	&--	&\multicolumn{2}{c|}{144}		&17,108	&3492	&\multicolumn{2}{c}{59}		 \\
	\cline{2-13}												
	&$gzip\_v0$	&1.0.7	&4.32	&1,468	&81	&--	&\multicolumn{2}{c|}{156}		&--	&--	&\multicolumn{2}{c}{--}		 \\
	&$gzip\_v1$	&1.1.2	&4.52	&1,490	&81	&--	&\multicolumn{2}{c|}{156}		&7,429	&639	&\multicolumn{2}{c}{8}		 \\
	&$gzip\_v2$	&1.2.2	&5.05	&1,752	&98	&--	&\multicolumn{2}{c|}{156}		&7,599	&659	&\multicolumn{2}{c}{8}		 \\
	&$gzip\_v3$	&1.2.3	&5.06	&1,610	&93	&--	&\multicolumn{2}{c|}{156}		&7,678	&547	&\multicolumn{2}{c}{7}		 \\
	&$gzip\_v4$	&1.2.4	&5.18	&1,663	&93	&--	&\multicolumn{2}{c|}{156}		&7,838	&548	&\multicolumn{2}{c}{7}		 \\
	&$gzip\_v5$	&1.3	&5.68	&1,733	&97	&--	&\multicolumn{2}{c|}{156}		&8,809	&210	&\multicolumn{2}{c}{7}		 \\
	\cline{2-13}												
	&$make\_v0$	&3.75	&17.46	&4,397	&181	&--	&\multicolumn{2}{c|}{111}		&--	&--	&\multicolumn{2}{c}{--}		 \\
	&$make\_v1$	&3.76.1	&18.57	&4,585	&181	&--	&\multicolumn{2}{c|}{111}		&36,262	&5,800	&\multicolumn{2}{c}{37}		 \\
	&$make\_v2$	&3.77	&19.66	&4,784	&190	&--	&\multicolumn{2}{c|}{111}		&38,183	&5,965	&\multicolumn{2}{c}{29}		 \\
	&$make\_v3$	&3.78.1	&20.46	&4,845	&216	&--	&\multicolumn{2}{c|}{111}		&42,281	&6,244	&\multicolumn{2}{c}{28}		 \\
	&$make\_v4$	&3.79	&23.13	&5,413	&239	&--	&\multicolumn{2}{c|}{111}		&48,546	&6,958	&\multicolumn{2}{c}{29}		 \\
	&$make\_v5$	&3.80	&23.40	&5,032	&268	&--	&\multicolumn{2}{c|}{111}		&47,310	&7,049	&\multicolumn{2}{c}{28}		 \\
	\cline{2-13}												
	&$sed\_v0$	&3.01	&7.79	&676	&66	&--	&\multicolumn{2}{c|}{324}		&--	&--	&\multicolumn{2}{c}{--}		 \\
	&$sed\_v1$	&3.02	&7.79	&712	&65	&--	&\multicolumn{2}{c|}{324}		&2,506	&1,009	&\multicolumn{2}{c}{16}		 \\
	&$sed\_v2$	&4.0.6	&18.55	&1,011	&65	&--	&\multicolumn{2}{c|}{324}		&5,947	&1,048	&\multicolumn{2}{c}{18}		 \\
	&$sed\_v3$	&4.0.8	&18.69	&1,017	&66	&--	&\multicolumn{2}{c|}{324}		&5,970	&450	&\multicolumn{2}{c}{18}		 \\
	&$sed\_v4$	&4.1.1	&21.74	&1,141	&70	&--	&\multicolumn{2}{c|}{324}		&6,578	&470	&\multicolumn{2}{c}{19}		 \\
	&$sed\_v$5	&4.2	&26.47	&1,412	&98	&--	&\multicolumn{2}{c|}{324}		&7,761	&628	&\multicolumn{2}{c}{22}		 \\
	 \hline												
 \end{tabular}}
\end{table*}

\begin{figure*}[!h]
  \centering
  \includegraphics[scale=0.5]{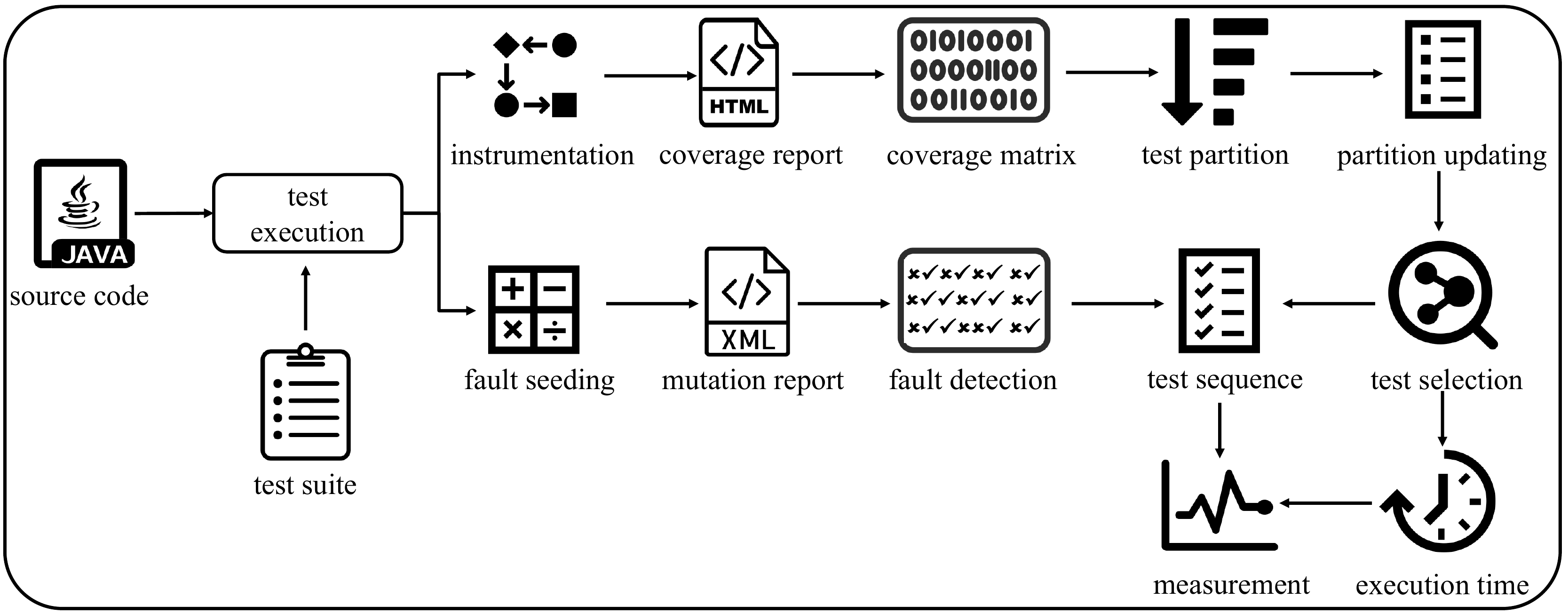}
  \caption{{\toolname}'s Framework.}
  \label{fig_tcp}
\end{figure*}

\subsection{Framework}

Figure \ref{fig_tcp} presents the overall experimental framework of the proposed technique.
(1) We collect the coverage information for the Java program using the \texttt{FaultTracer} tool \cite{2012Zhang,2013Zhanga}, which uses on-the-fly bytecode instrumentation without any modification of the target program based on the ASM bytecode manipulation and analysis framework \cite{asm}.
For C program, there are six versions of each program $P$: $P_{V0}$, $P_{V1}$, $P_{V2}$, $P_{V3}$, $P_{V4}$, and $P_{V5}$.
Version $P_{V0}$ is compiled using \texttt{gcc} 5.4.0 \cite{gcc}, and then the coverage information is obtained using the \texttt{gcov} tool~\cite{gcov}.
(2) After collecting the code coverage information, we implement all TCP techniques in Java, and apply them to each program version under study.
Specifically, {\toolname} first divides test cases into different partitions and updates test cases in the highest partition.
If there exist an updated test case satisfies the selection criteria, the test case will be added to the prioritized test sequence.
Because the approaches contain randomness, each execution is repeated 1,000 times independently.
This results in, for each testing scenario, 1,000 test sequences for each TCP technique.
(3) To evaluate the fault detection rate, we construct the faulty programs by mutation faults.
Specifically, we generate mutants by seeding all mutation operators (presented in Table \ref{TAB:Mutators}) and consider each mutant as a faulty program with only one mutation fault.
We then execute all test cases against each faulty program and remove the mutants that any test case cannot kill.
(4) Besides, we calculate the APFD values and prioritization time for all test sequences based on the record information (e.g., the mutation detected results and time cost of each TCP technique)
(5) To further test whether there is a statistically significant difference between {\toolname} and other TCP techniques, we perform the unpaired two-tailed Wilcoxon-Mann-Whitney test, at a significance level of $5\%$, following previously reported guidelines for inferential statistical analysis involving randomized algorithms~\cite{2014Arcuri,2015Gligoric}.
To identify which technique is better, we also calculate the effect size, measured by the non-parametric Vargha and Delaney effect size measure~\cite{2000Vargha}, $\hat{\textrm{A}}_{12}$, where $\hat{\textrm{A}}_{12} (X,Y)$ gives probability that the technique $X$ is better than technique $Y$.
The statistical analyses are performed using R language \cite{Rstatis}.

\subsection{Experimental Setup}
The experiments are conducted on a Linux 5.15.0-25-generic cloud server with eight virtual cores of Intel(R) Xeon(R) Silver 4116 CPU (2.10 GHz) and 32 GBs of virtual RAM.

\section{Results and analysis}
\label{sec:re&an}

This section presents the experimental results to answer the research questions.
We investigate the effectiveness of {\toolname} to answer RQ1, and perform impact analysis to investigate the influences caused by the code coverage granularity to answer RQ2. 
Besides, we also perform analysis to investigate the influences caused by and test case granularity on {\toolname} to answer RQ3. 
Finally, we analyze the time cost of {\toolname} to answer RQ4. 

To answer RQ1 to RQ3, Figures \ref{FIG:apfd-c} to \ref{FIG:SBM} present box plots of the distribution of the APFD values achieved over 1,000 independent runs.
Each box plot shows the mean (square in the box), median (line in the box), and upper and lower quartiles (25th and 75th percentile) of the APFD values for all the TCP techniques.
Statistical analyses are also provided in Tables \ref{TAB:apfd} to \ref{TAB:SBM} for each pairwise APFD comparison between {\toolname} and the other TCP techniques.
For ease of illustration, we denote the mentioned TCP techniques as $TCP_{tot}$, $TCP_{add}$, $TCP_{lexi}$, $TCP_{unif}$, $TCP_{art}$, $TCP_{search}$ and $TCP_{\MakeLowercase{\toolname}}$, respectively.
For example, for a comparison between two methods
$TCP_{\MakeLowercase{\toolname}}$ and $M$,
where $M \in \{TCP_{tot}, TCP_{add}, TCP_{lexi}, TCP_{unif}, TCP_{art}, TCP_{search}\}$,
the symbol \ding{52} means that $TCP_{\MakeLowercase{\toolname}}$ is better
($p$-value is less than 0.05, and the effect size $\hat{\textrm{A}}_{12}(TCP_{\MakeLowercase{\toolname}},M)$ is greater than 0.50);
the symbol \ding{54} means that $M$ is better
(the $p$-value is less than 0.05, and $\hat{\textrm{A}}_{12}(TCP_{\MakeLowercase{\toolname}},M)$ is less than 0.50);
and the symbol \ding{109} means that there is no statistically significant difference between them (i.e., the $p$-value is greater than 0.05).

To answer RQ4, Table \ref{TAB:time} provides comparisons of the execution times for the different TCP techniques.

\subsection{RQ1: Effectiveness of {\toolname}}

\begin{figure*}[htbp]
\graphicspath{{graphs/}}
\centering
    \subfigure[statement coverage]
    {
        \includegraphics[width=0.31\textwidth]{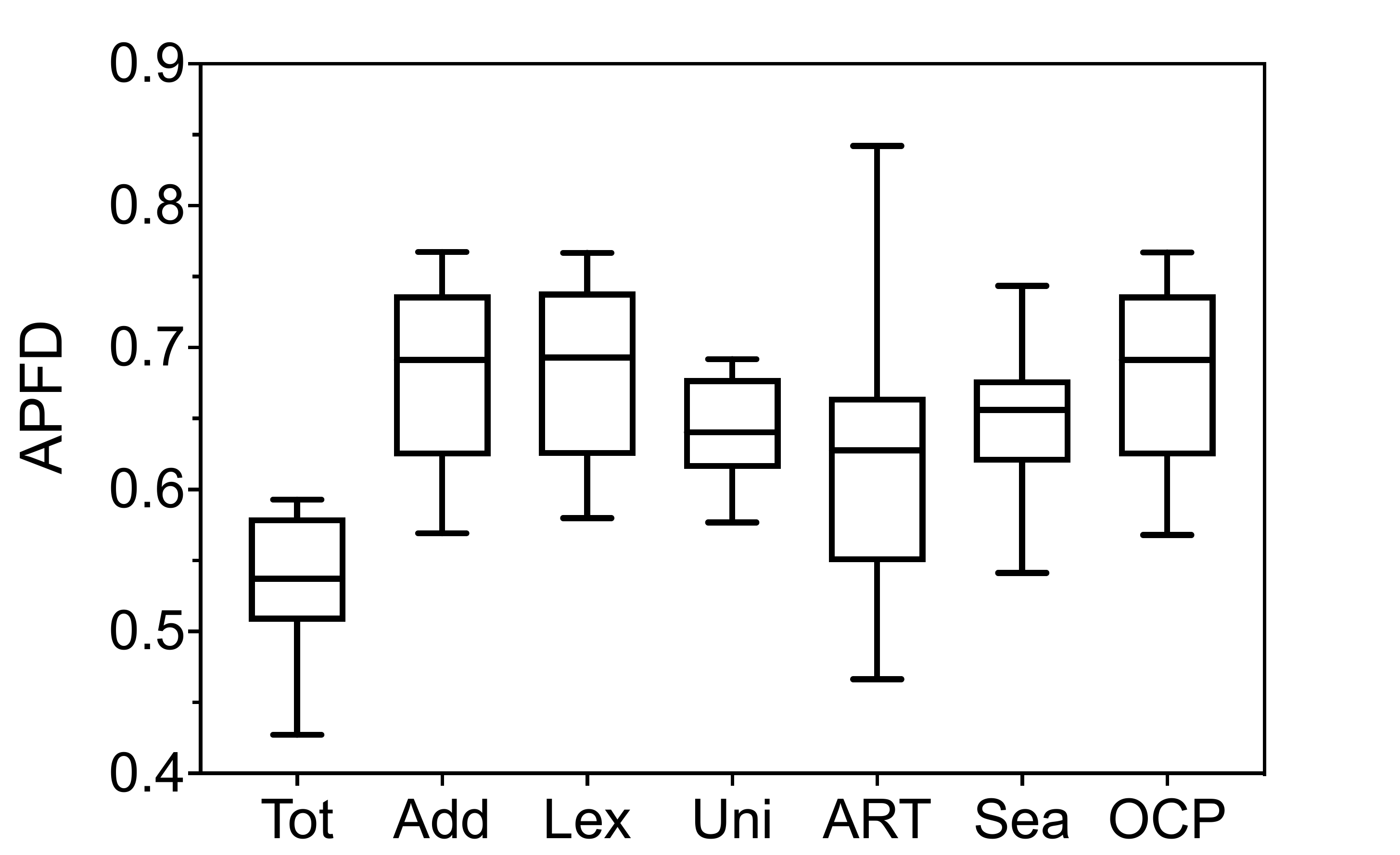}
        \label{apfd_flex_sc}
    }
    \subfigure[branch coverage]
    {
        \includegraphics[width=0.31\textwidth]{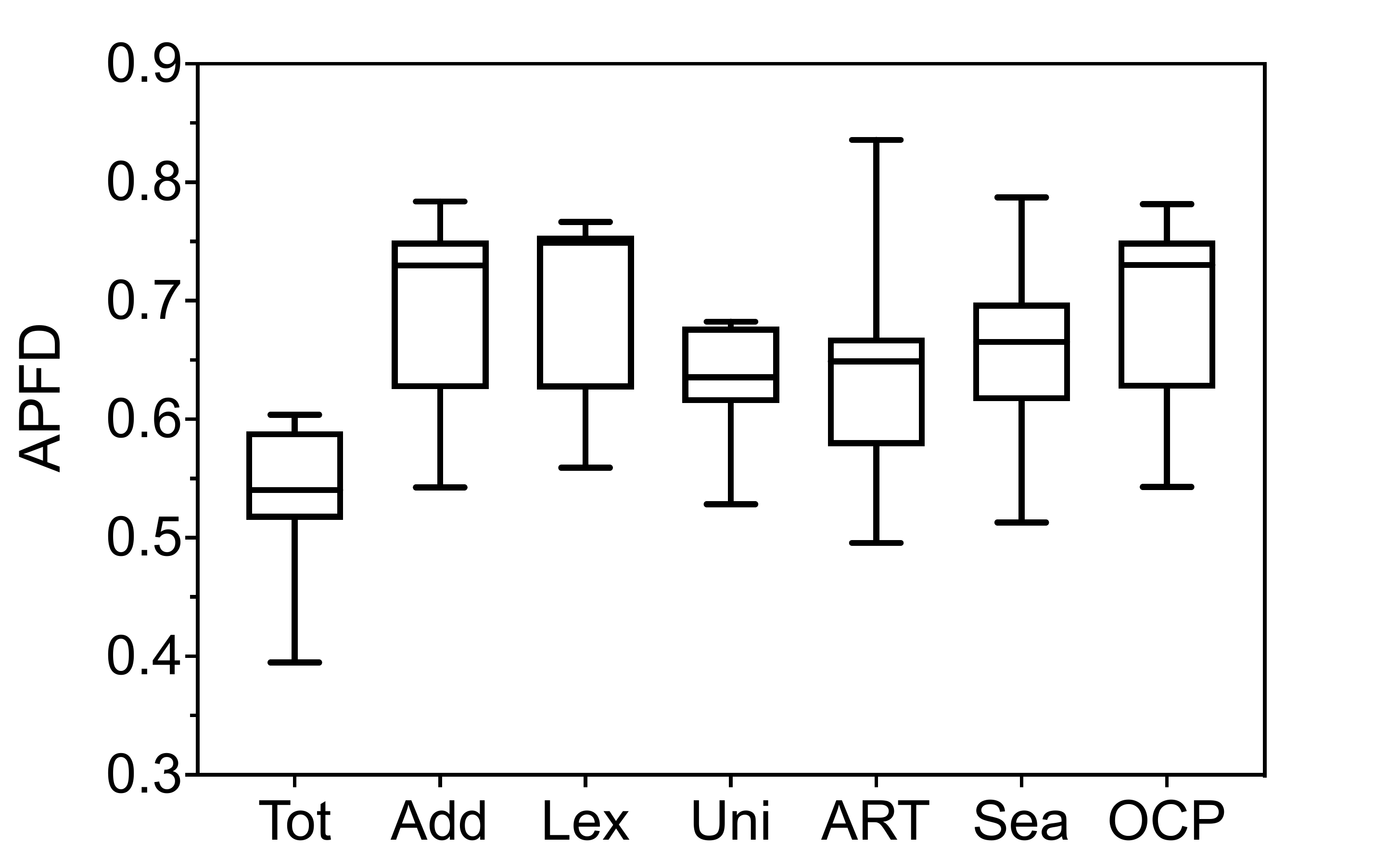}
        \label{apfd_flex_bc}
    }
        \subfigure[method coverage]
    {
        \includegraphics[width=0.31\textwidth]{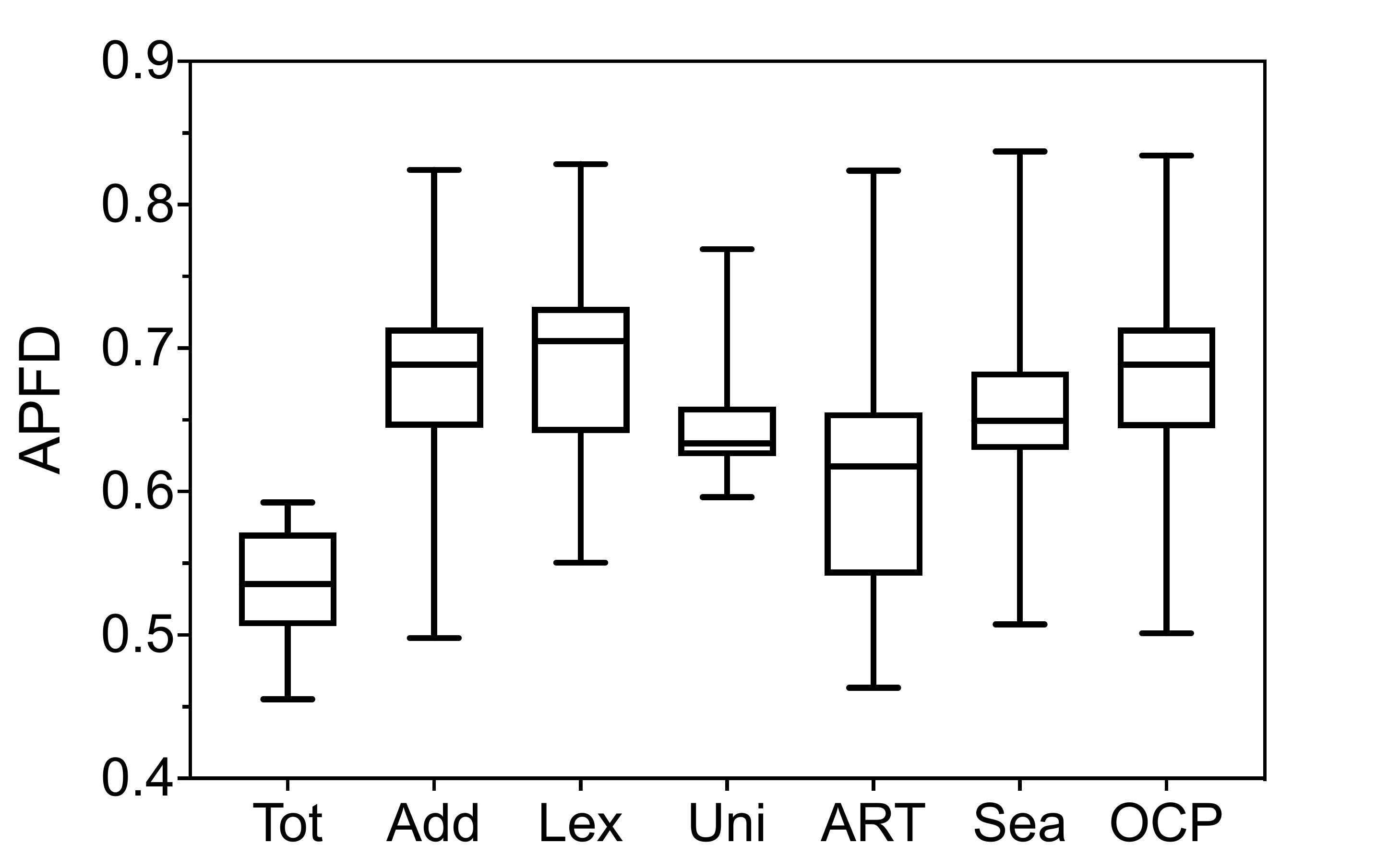}
        \label{apfd_flex_mc}
    }
    \caption{\textbf{APFD} results for Java programs at test-method granularity}
    \label{FIG:apfd-java-m}
\end{figure*}

\begin{figure*}[htbp]
\graphicspath{{graphs/}}
\centering
    \subfigure[statement coverage]
    {
        \includegraphics[width=0.31\textwidth]{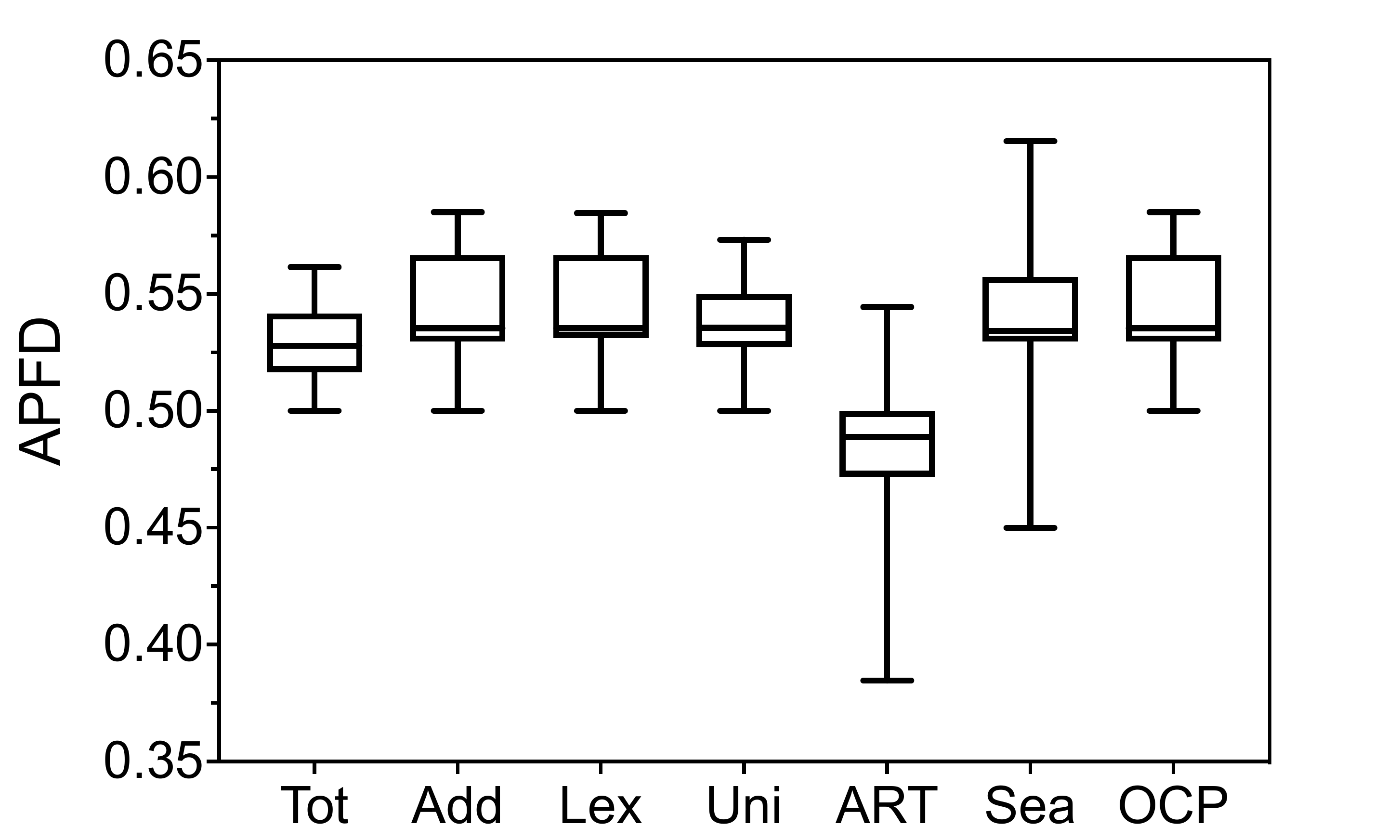}
        \label{apfd_flex_sc}
    }
    \subfigure[branch coverage]
    {
        \includegraphics[width=0.31\textwidth]{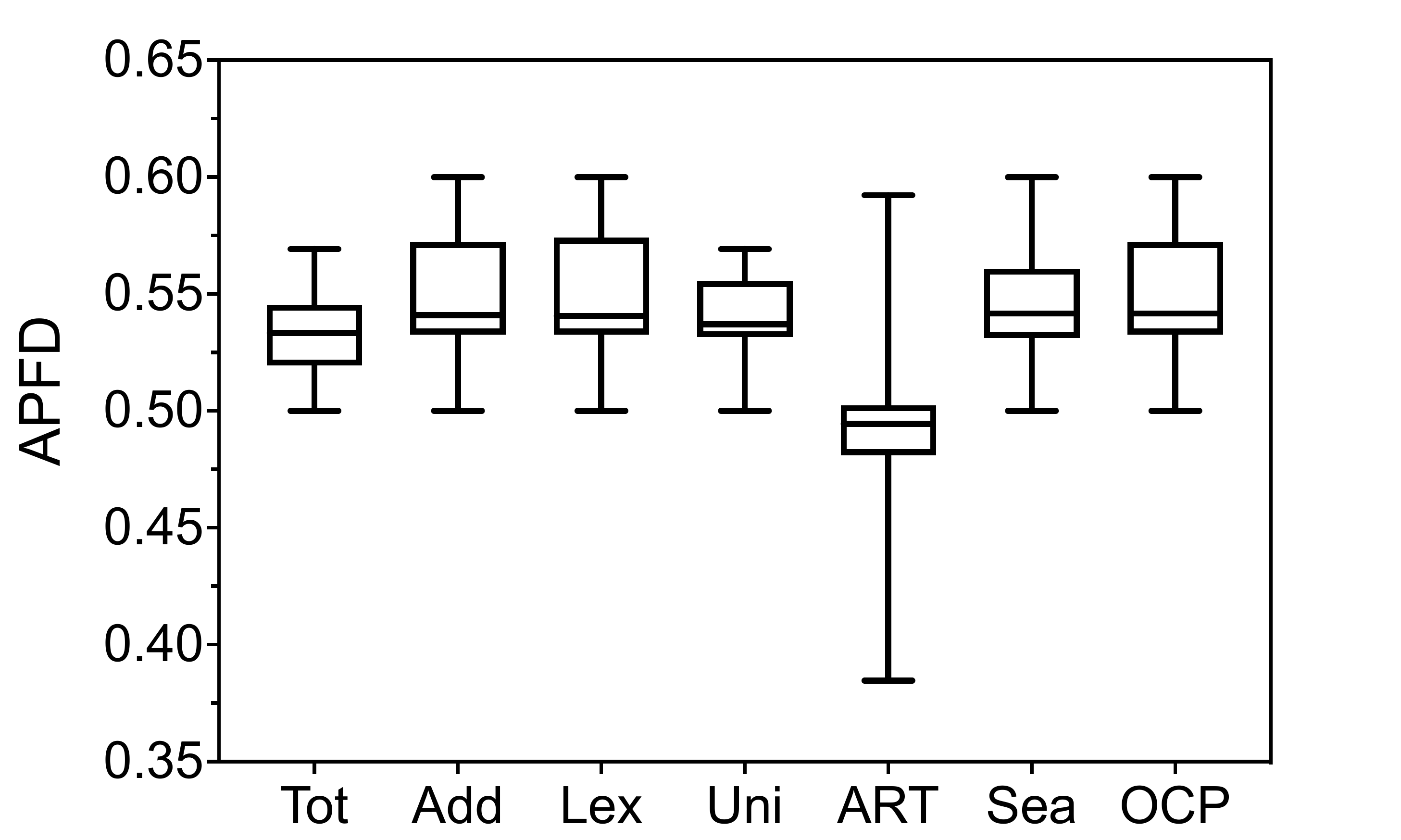}
        \label{apfd_flex_bc}
    }
        \subfigure[method coverage]
    {
        \includegraphics[width=0.31\textwidth]{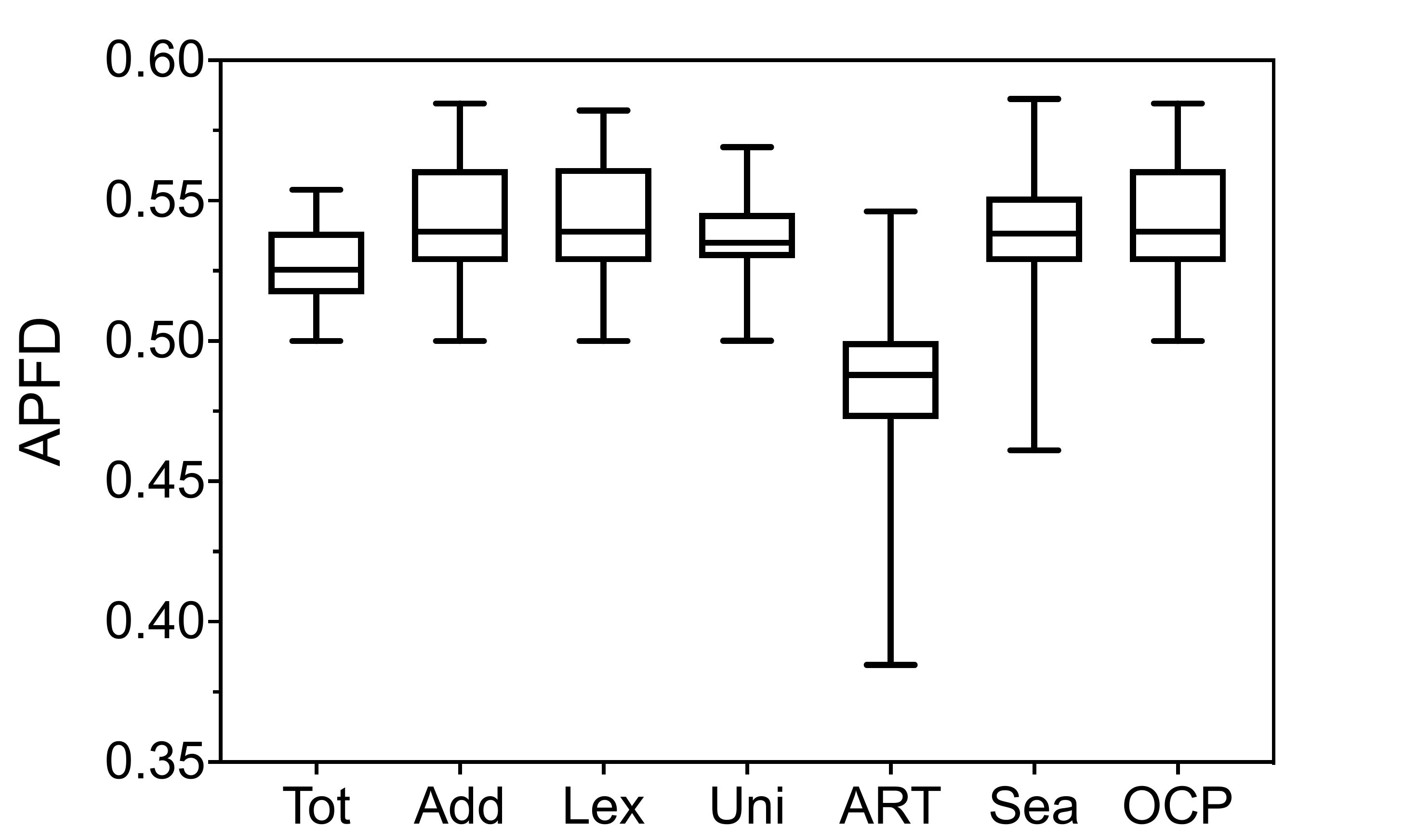}
        \label{apfd_flex_mc}
    }
    \caption{\textbf{APFD} results for \textbf{Java programs} at test-class granularity}
    \label{FIG:apfd-java-c}
\end{figure*}

In this section, we evaluate the effectiveness of different TCP techniques by fault detection rate.
We provide the APFD results for {\toolname} with different code coverage criteria and test case granularities.
Figures \ref{FIG:apfd-c} to \ref{FIG:apfd-java-c} show the APFD results for the C programs, the Java programs at the test-method granularity and the test-method granularity, respectively.
Each sub-figure in these figures has the seven TCP techniques across the $x$-axis, and corresponding to the APFD values on the $y$-axis.
Table \ref{TAB:apfd} presents the corresponding statistical comparisons.
Each row denotes the statistical results for the corresponding program under different coverage criteria.
Column "C Programs", "Java-M Programs", "Java-C Programs" and"All Programs" are calculated based on all APFD values for C programs, Java programs at the test-method granularity, Java programs at the test-class granularity and all programs.

\subsubsection{C Subject Programs}

\begin{figure*}[!t]
\graphicspath{{graphs/}}
\centering
    \subfigure[statement coverage]
    {
        \includegraphics[width=0.31\textwidth]{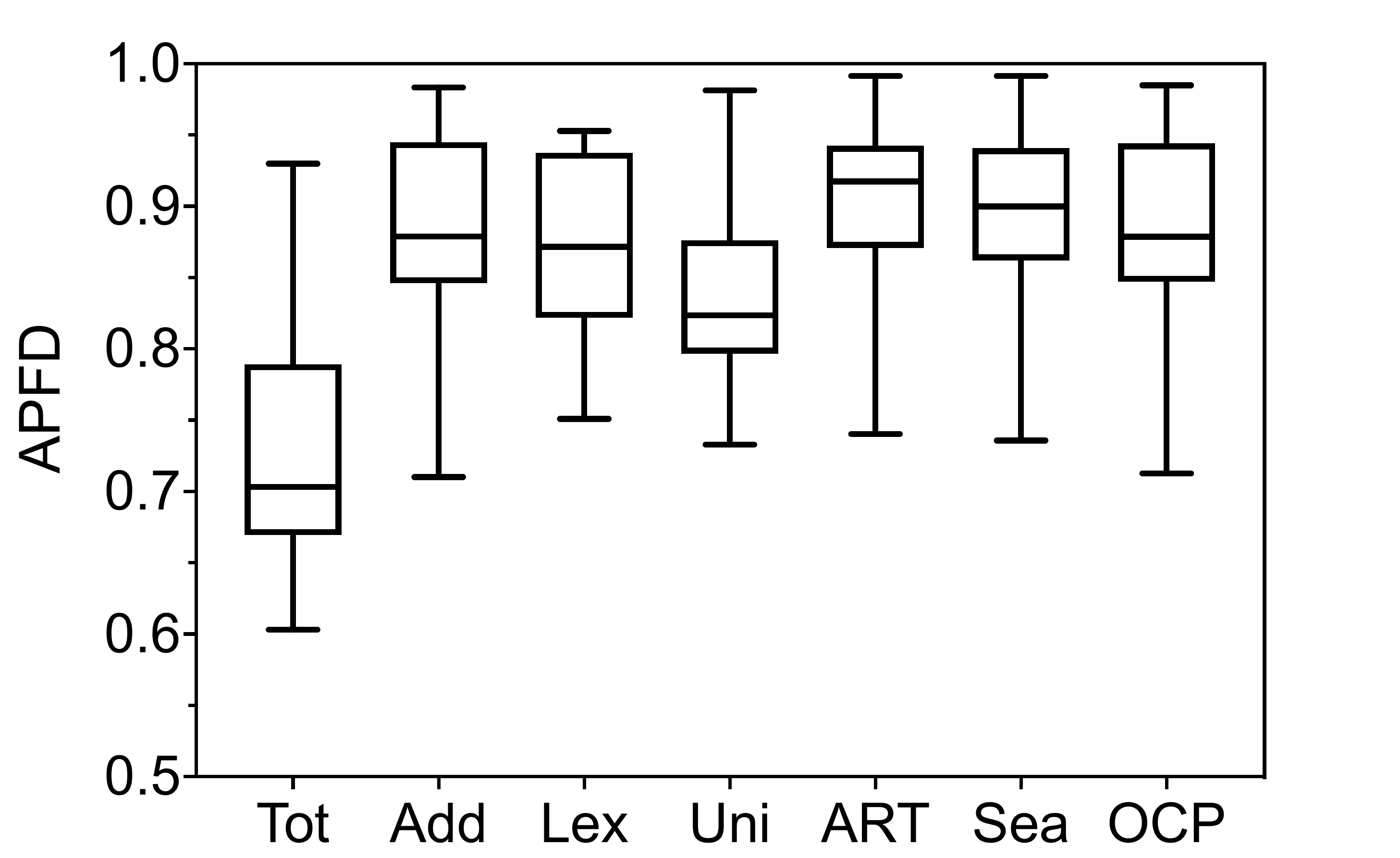}
        \label{apfd_flex_sc}
    }
    \subfigure[branch coverage]
    {
        \includegraphics[width=0.31\textwidth]{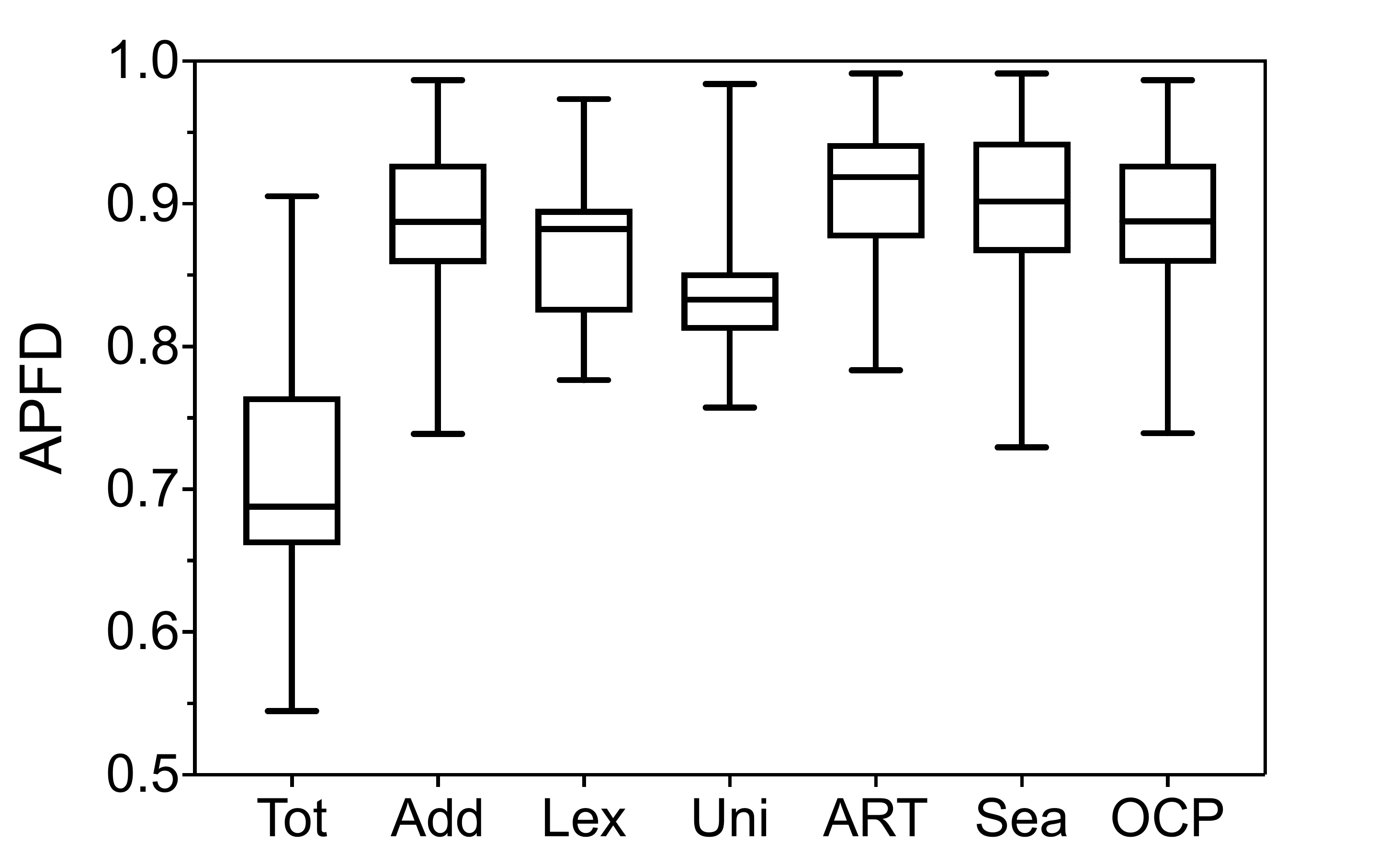}
        \label{apfd_flex_bc}
    }
        \subfigure[method coverage]
    {
        \includegraphics[width=0.31\textwidth]{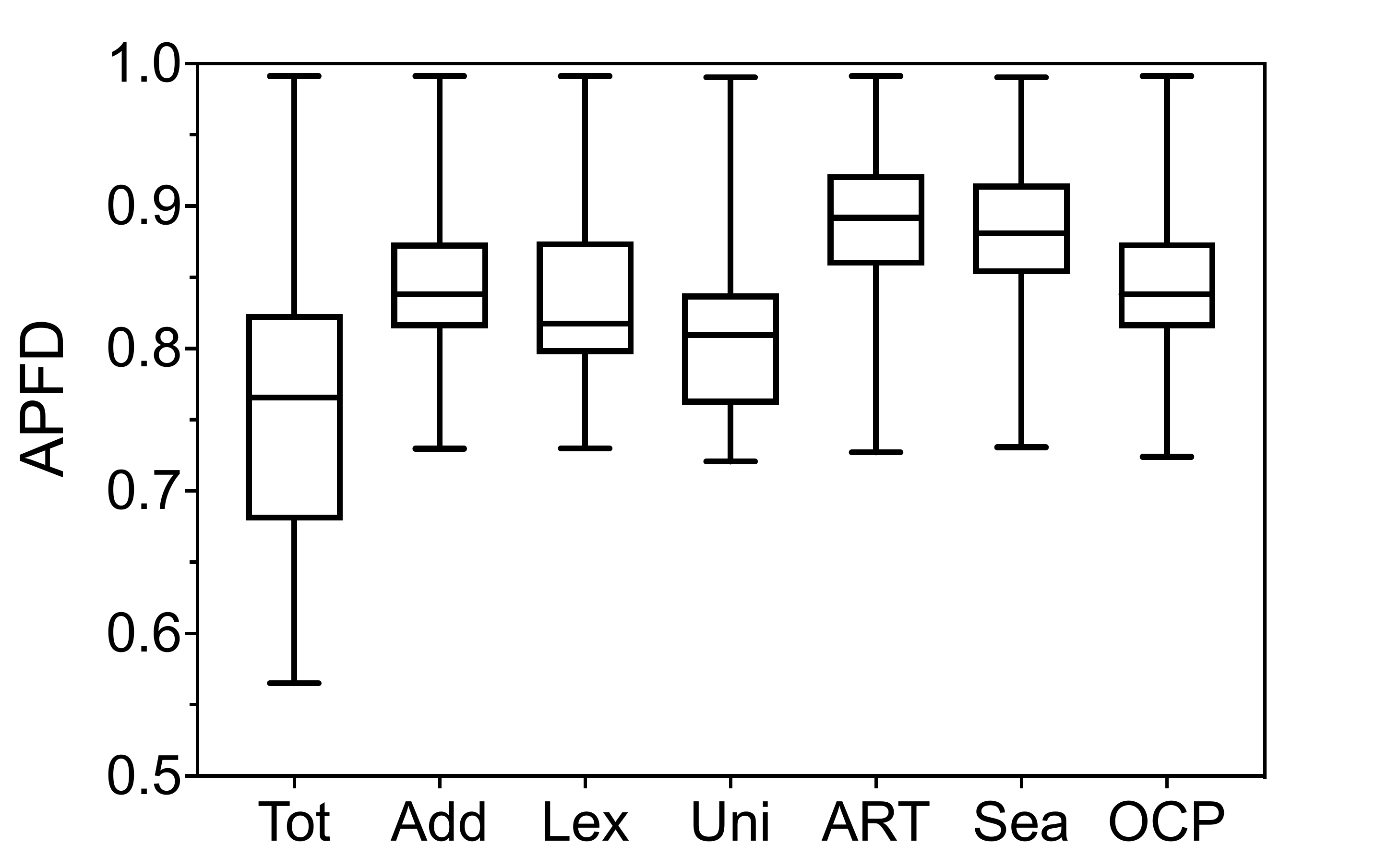}
        \label{apfd_flex_mc}
    }
    \caption{\textbf{APFD} results for \textbf{C programs}}
    \label{FIG:apfd-c}
\end{figure*}

Based the results on Figure \ref{FIG:apfd-c} and Table \ref{TAB:apfd}, we make the following observations:

When comparing $TCP_{\MakeLowercase{\toolname}}$ with the \textit{greedy-based} strategies,
our proposed $TCP_{\MakeLowercase{\toolname}}$ approach has much better performance than $TCP_{tot}$, $TCP_{add}$, $TCP_{unify}$ and $TCP_{lexi}$ for all programs and code coverage granularities, except for $make$ with branch coverage (for which $TCP_{\MakeLowercase{\toolname}}$ has very similar, or better performance).
The maximum difference in mean and median APFD values between $TCP_{\MakeLowercase{\toolname}}$ and $TCP_{tot}$ is more than 40\%,
while between $TCP_{\MakeLowercase{\toolname}}$ and $TCP_{add}$, it is about 10\%.

Our proposed technique $TCP_{\MakeLowercase{\toolname}}$ has similar or better APFD performance than $TCP_{art}$ and $TCP_{search}$ for some subject programs
(e.g.,  $flex$ and $gzip$), with all code coverage granularities,
but has slightly worse performance for some others
(e.g., $make$ and $sed$).
However, the difference in mean and median APFD values between $TCP_{\MakeLowercase{\toolname}}$ and $TCP_{art}$ or $TCP_{search}$
is less than 5\%.

Furthermore, the statistical results support the box plot observations.
All $p$-values for the comparisons between $TCP_{\MakeLowercase{\toolname}}$ and the \textit{greedy-based} strategies (i.e., $TCP_{tot}$ or $TCP_{add}$) are less than 0.05 (except for $make$ with branch coverage),
indicating that their APFD scores are significantly different.
The $\hat{\textrm{A}}_{12}$ values are also much greater than 0.50, ranging from 0.51 to 1.00.
However, although all $p$-values between $TCP_{\MakeLowercase{\toolname}}$ and $TCP_{art}$ or $\textit{TCP}_\textit{search}$ are also less than 0.05,
their $\hat{\textrm{A}}_{12}$ values are much greater than 0.50 in some cases, but much less than 0.50 in others.
Nevertheless, considering all the C programs,
not only does $TCP_{\MakeLowercase{\toolname}}$ have significantly different APFD values to the other six TCP techniques,
but it also has better performances overall (except for $TCP_{art}$ and $TCP_{search}$).

\subsubsection{Java Programs at Test-Method Granularity}

Based on Figure \ref{FIG:apfd-java-c} and Table \ref{TAB:apfd}, we have the following observations:

Compared with the \textit{greedy-based} strategies, $TCP_{\MakeLowercase{\toolname}}$ performs much better than $TCP_{tot}$, regardless of subject program and code coverage granularity, with the maximum mean and median APFD differences reaching about 12\%.
$TCP_{\MakeLowercase{\toolname}}$ also has very similar performance to $TCP_{add}$, with the mean and median APFD differences approximately equal to 1\%.
However, none of the other two techniques ($TCP_{lexi}$ and $TCP_{unify}$) is either always better or always worse than $TCP_{\MakeLowercase{\toolname}}$, with $TCP_{\MakeLowercase{\toolname}}$ sometimes performing better for some programs, and sometimes worse.

$TCP_{\MakeLowercase{\toolname}}$ also performs better than $\textit{TCP}_\textit{art}$ and $\textit{TCP}_\textit{search}$ at most cases (except $jtopas$ at statement and branch coverage)
There is a statistically significant difference between $TCP_{\MakeLowercase{\toolname}}$ and $\textit{TCP}_\textit{art}$, which supports the above observations.

Furthermore, the statistical results support the box plot observations.
Considering all Java programs, 
$TCP_{\MakeLowercase{\toolname}}$ performs better than $TCP_{tot}$, $TCP_{art}$ and $TCP_{search}$, as most $p$-values are less than 0.05, and the relevant effect size $\hat{\textrm{A}}_{12}$ ranges from 0.58 to 0.98.
However, {\toolname} has very a similar (or slightly worse) performance to $TCP_{add}$, with $\hat{\textrm{A}}_{12}$ values of either 0.48 or 0.50.

\begin{table*}[t]
\tiny
\centering
 \caption{Statistical \textbf{effectiveness}  comparisons of \textbf{APFD} for \textbf{all programs}.}
  \label{TAB:apfd}
    \setlength{\tabcolsep}{0.8mm}
    {
    \begin{tabular}{m{1cm}<{\centering}|c|cccccc|cccccc|cccccc}
     \hline
       \multicolumn{2}{c|}{\multirow{2}{*}{\textbf{Program Name}}}  &\multicolumn{6}{c|}{\textbf{Statement Coverage}} &\multicolumn{6}{c|}{\textbf{Branch Coverage}} &\multicolumn{6}{c}{\textbf{Method Coverage}} \\
        \cline{3-20}

         \multicolumn{2}{c|}{} & $\textit{TCP}_\textit{tot}$ &$\textit{TCP}_\textit{add}$ &$\textit{TCP}_\textit{unify}$ &$\textit{TCP}_\textit{lexi}$ &$\textit{TCP}_\textit{art}$ &$\textit{TCP}_\textit{search}$  &
          $\textit{TCP}_\textit{tot}$ &$\textit{TCP}_\textit{add}$ &$\textit{TCP}_\textit{unify}$ &$\textit{TCP}_\textit{lexi}$ &$\textit{TCP}_\textit{art}$ &$\textit{TCP}_\textit{search}$ &
          $\textit{TCP}_\textit{tot}$ &$\textit{TCP}_\textit{add}$ &$\textit{TCP}_\textit{unify}$ &$\textit{TCP}_\textit{lexi}$ &$\textit{TCP}_\textit{art}$ &$\textit{TCP}_\textit{search}$\\
            \hline



$gnu\_flex$ & \multirow{5}*{\textbf{--}} & \ding{52} (1.00) & \ding{52} (0.52) & \ding{52} (0.70) & \ding{52} (0.87) & \ding{52} (0.86) & \ding{52} (0.73) & \ding{52} (1.00) & \ding{52} (0.66) & \ding{54} (0.22) & \ding{52} (0.90) & \ding{52} (0.72) & \ding{52} (0.57) & \ding{52} (1.00) & \ding{52} (0.70) & \ding{52} (0.71) & \ding{52} (0.85) & \ding{54} (0.28) & \ding{54} (0.47) \\
$gnu\_make$ &  & \ding{52} (0.85) & \ding{52} (0.61) & \ding{52} (0.62) & \ding{52} (0.82) & \ding{54} (0.30) & \ding{54} (0.29) & \ding{52} (1.00) & \ding{109} (0.51) & \ding{52} (0.77) & \ding{52} (0.87) & \ding{54} (0.28) & \ding{54} (0.42) & \ding{52} (0.58) & \ding{52} (0.68) & \ding{52} (0.68) & \ding{52} (0.58) & \ding{54} (0.22) & \ding{54} (0.24) \\
$gnu\_grep$ &  & \ding{52} (1.00) & \ding{52} (0.57) & \ding{52} (0.79) & \ding{52} (1.00) & \ding{54} (0.48) & \ding{52} (0.64) & \ding{52} (1.00) & \ding{52} (0.54) & \ding{52} (0.67) & \ding{52} (1.00) & \ding{109} (0.50) & \ding{52} (0.66) & \ding{52} (1.00) & \ding{52} (0.71) & \ding{52} (0.99) & \ding{52} (1.00) & \ding{54} (0.21) & \ding{54} (0.25) \\
$gnu\_gzip$ &  & \ding{52} (0.80) & \ding{52} (0.66) & \ding{52} (0.78) & \ding{52} (0.58) & \ding{54} (0.43) & \ding{54} (0.36) & \ding{52} (0.84) & \ding{52} (0.70) & \ding{52} (0.83) & \ding{109} (0.49) & \ding{54} (0.37) & \ding{54} (0.32) & \ding{52} (0.54) & \ding{52} (0.53) & \ding{52} (0.54) & \ding{52} (0.53) & \ding{54} (0.47) & \ding{54} (0.47) \\
$gnu\_sed$ &  & \ding{52} (1.00) & \ding{52} (0.54) & \ding{52} (0.64) & \ding{52} (1.00) & \ding{54} (0.15) & \ding{54} (0.37) & \ding{52} (1.00) & \ding{52} (0.55) & \ding{52} (0.64) & \ding{52} (1.00) & \ding{54} (0.14) & \ding{54} (0.37) & \ding{52} (1.00) & \ding{52} (0.90) & \ding{52} (0.97) & \ding{52} (0.97) & \ding{54} (0.22) & \ding{54} (0.29) \\
\hline
\multicolumn{2}{c|}{$\textit{\textbf{C Programs}}$} & \ding{52} (0.91) & \ding{52} (0.54) & \ding{52} (0.60) & \ding{52} (0.74) & \ding{54} (0.44) & \ding{54} (0.47) & \ding{52} (0.93) & \ding{52} (0.55) & \ding{52} (0.60) & \ding{52} (0.77) & \ding{54} (0.43) & \ding{54} (0.47) & \ding{52} (0.81) & \ding{52} (0.61) & \ding{52} (0.66) & \ding{52} (0.74) & \ding{54} (0.37) & \ding{54} (0.41) \\
\hline

$ant$ & \multirow{4}{*}{\rotatebox{90}{\textbf{method}}} & \ding{52} (1.00) & \ding{109} (0.50) & \ding{54} (0.42) & \ding{52} (1.00) & \ding{52} (1.00) & \ding{52} (1.00) & \ding{52} (1.00) & \ding{109} (0.50) & \ding{54} (0.40) & \ding{52} (1.00) & \ding{52} (1.00) & \ding{52} (1.00) & \ding{52} (1.00) & \ding{109} (0.50) & \ding{54} (0.22) & \ding{52} (0.98) & \ding{52} (0.96) & \ding{52} (0.95) \\
$jtopas$ &  & \ding{52} (1.00) & \ding{109} (0.50) & \ding{54} (0.36) & \ding{54} (0.39) & \ding{54} (0.00) & \ding{54} (0.44) & \ding{52} (1.00) & \ding{109} (0.50) & \ding{54} (0.38) & \ding{52} (1.00) & \ding{54} (0.00) & \ding{54} (0.44) & \ding{52} (1.00) & \ding{109} (0.50) & \ding{54} (0.03) & \ding{52} (0.66) & \ding{52} (0.56) & \ding{109} (0.50) \\
$jmeter$ &  & \ding{52} (1.00) & \ding{109} (0.50) & \ding{109} (0.50) & \ding{54} (0.38) & \ding{52} (1.00) & \ding{52} (0.62) & \ding{52} (1.00) & \ding{54} (0.48) & \ding{52} (0.61) & \ding{109} (0.51) & \ding{52} (1.00) & \ding{52} (0.69) & \ding{52} (0.86) & \ding{109} (0.50) & \ding{54} (0.47) & \ding{54} (0.45) & \ding{52} (0.98) & \ding{52} (0.58) \\
$xmlsec$ &  & \ding{52} (1.00) & \ding{109} (0.50) & \ding{54} (0.39) & \ding{52} (1.00) & \ding{52} (1.00) & \ding{109} (0.49) & \ding{52} (1.00) & \ding{109} (0.49) & \ding{54} (0.13) & \ding{52} (1.00) & \ding{52} (1.00) & \ding{52} (0.77) & \ding{52} (1.00) & \ding{109} (0.50) & \ding{54} (0.36) & \ding{52} (0.99) & \ding{52} (1.00) & \ding{52} (0.55) \\
\hline
\multicolumn{2}{c|}{$\textit{\textbf{Java-M Programs}}$} & \ding{52} (0.98) & \ding{109} (0.50) & \ding{54} (0.48) & \ding{52} (0.67) & \ding{52} (0.70) & \ding{52} (0.65) & \ding{52} (0.94) & \ding{109} (0.50) & \ding{54} (0.45) & \ding{52} (0.70) & \ding{52} (0.64) & \ding{52} (0.66) & \ding{52} (0.98) & \ding{109} (0.50) & \ding{54} (0.39) & \ding{52} (0.71) & \ding{52} (0.77) & \ding{52} (0.64) \\
\hline

$ant$ & \multirow{4}{*}{\rotatebox{90}{\textbf{class}}} & \ding{52} (0.83) & \ding{109} (0.50) & \ding{52} (0.52) & \ding{52} (0.70) & \ding{52} (0.99) & \ding{52} (0.60) & \ding{52} (0.89) & \ding{109} (0.50) & \ding{52} (0.51) & \ding{52} (0.69) & \ding{52} (1.00) & \ding{52} (0.54) & \ding{52} (0.94) & \ding{109} (0.50) & \ding{54} (0.48) & \ding{52} (0.71) & \ding{52} (1.00) & \ding{52} (0.57) \\
$jtopas$ &  & \ding{109} (0.50) & \ding{109} (0.50) & \ding{109} (0.50) & \ding{109} (0.50) & \ding{52} (0.93) & \ding{109} (0.50) & \ding{109} (0.50) & \ding{109} (0.50) & \ding{109} (0.50) & \ding{109} (0.50) & \ding{52} (0.91) & \ding{109} (0.50) & \ding{54} (0.33) & \ding{109} (0.50) & \ding{109} (0.50) & \ding{54} (0.33) & \ding{52} (0.94) & \ding{109} (0.50) \\
$jmeter$ &  & \ding{54} (0.36) & \ding{109} (0.50) & \ding{54} (0.46) & \ding{54} (0.36) & \ding{52} (1.00) & \ding{109} (0.50) & \ding{54} (0.40) & \ding{54} (0.47) & \ding{52} (0.52) & \ding{54} (0.40) & \ding{52} (1.00) & \ding{52} (0.52) & \ding{54} (0.31) & \ding{109} (0.50) & \ding{54} (0.48) & \ding{54} (0.32) & \ding{52} (1.00) & \ding{109} (0.49) \\
$xmlsec$ &  & \ding{52} (1.00) & \ding{109} (0.50) & \ding{109} (0.50) & \ding{52} (1.00) & \ding{52} (1.00) & \ding{109} (0.50) & \ding{52} (0.97) & \ding{54} (0.48) & \ding{54} (0.40) & \ding{52} (0.96) & \ding{52} (1.00) & \ding{109} (0.50) & \ding{52} (1.00) & \ding{109} (0.51) & \ding{52} (0.60) & \ding{52} (1.00) & \ding{52} (1.00) & \ding{52} (0.58) \\
\hline
\multicolumn{2}{c|}{$\textit{\textbf{Java-C Programs}}$} & \ding{52} (0.63) & \ding{109} (0.50) & \ding{109} (0.50) & \ding{52} (0.55) & \ding{52} (0.98) & \ding{52} (0.53) & \ding{52} (0.65) & \ding{109} (0.50) & \ding{109} (0.50) & \ding{52} (0.56) & \ding{52} (0.97) & \ding{52} (0.51) & \ding{52} (0.67) & \ding{109} (0.50) & \ding{109} (0.50) & \ding{52} (0.58) & \ding{52} (0.98) & \ding{52} (0.52) \\
\hline

\multicolumn{2}{c|}{$\textit{\textbf{All Programs}}$} & \ding{52} (0.71) & \ding{52} (0.51) & \ding{52} (0.51) & \ding{52} (0.56) & \ding{52} (0.57) & \ding{52} (0.51) & \ding{52} (0.73) & \ding{52} (0.51) & \ding{52} (0.51) & \ding{52} (0.57) & \ding{52} (0.55) & \ding{52} (0.51) & \ding{52} (0.69) & \ding{52} (0.52) & \ding{52} (0.51) & \ding{52} (0.57) & \ding{52} (0.57) & \ding{109} (0.50) \\

\hline

  \end{tabular}}
\end{table*}

\subsubsection{Java Programs at Test-Class Granularity}

Based on Figure \ref{FIG:apfd-java-m} and Table \ref{TAB:apfd}, we have the following observations:

{\toolname} achieves higher mean and median APFD values than $TCP_{tot}$ for most cases, except $jmeter$.
{\toolname} has a very similar performance to $TCP_{add}$, with their mean and median $APFD$ differences at around 1\%.
{\toolname} has a competitive performance with $TCP_{unify}$ and $TCP_{lexi}$ for all programs with different code coverage granularities.
{\toolname} achieves much higher mean and median APFD values than $TCP_{art}$ for most cases, for all programs with all code coverage granularities, with the maximum differences reaching approximately 10\%.
Other than for a few cases (e.g., $jtopas$), {\toolname} usually has better performance than $TCP_{search}$.

Furthermore, the statistical analysis supports the above box plots observations.
Considering all Java programs together,
{\toolname} proforms better than $TCP_{tot}$, $TCP_{unify}$, $TCP_{search}$, $TCP_{art}$, and $TCP_{search}$ on the whole.
Most $p$-values are less than 0.05, indicating that their differences are significant;
and the effect size $\hat{\textrm{A}}_{12}$ values range up to 1.00, which means that $TCP_{\MakeLowercase{\toolname}}$ is better than the other five TCP techniques.
Finally, while the $p$-values for comparisons between $TCP_{\MakeLowercase{\toolname}}$ and $TCP_{add}$ are less than 0.05 (which means that the differences are insignificant), the $\hat{\textrm{A}}_{12}$ values range from 0.49 to 0.51, indicating that they are very similar.

\finding{1}{
Overall, our analysis on the fault detection effectiveness that
(1) For C programs, {\toolname} has significantly better performance than all \textit{greedy-based} strategies and maintaining the comparable performance with $TCP_{art}$ and $TCP_{sea}$.
(2) For Java programs at test-method granularity, {\toolname} has better performance than $TCP_{tot}$, $TCP_{art}$ and $TCP_{sea}$, while has similar performance  with $TCP_{add}$, $TCP_{unify}$ and $TCP_{lexi}$.
(3) For Java programs at test-class granularity, {\toolname} has better or similar performance with $TCP_{add}$, $TCP_{art}$ and $TCP_{sea}$ for all programs, while has comparable performance  with $TCP_{tot}$, $TCP_{unify}$ and $TCP_{lexi}$.
}

\begin{figure}[!b]
\graphicspath{{graphs/}}
\centering
    \includegraphics[width=0.47\textwidth]{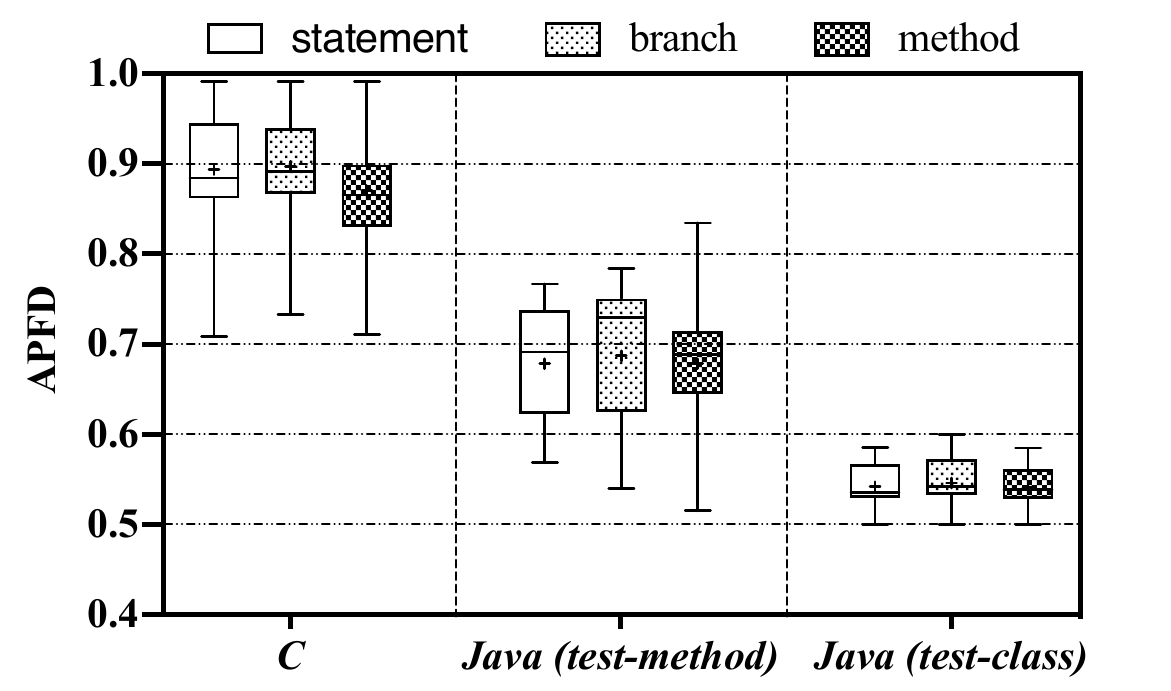}
    \caption{\textbf{Effectiveness:} \textbf{APFD} results with different code coverage and test case granularities for all programs}
    \label{FIG:SBM}
\end{figure}

\begin{table*}[htbp]
\scriptsize
\centering
 \caption{ Statistical \textbf{effectiveness} comparisons of \textbf{APFD} between different coverage granularities for {\toolname}.
 }
  \label{TAB:SBM}
  \setlength{\tabcolsep}{1.3mm}{
    \begin{tabular}{|c|c|rrr|rrr|rrr|}
     \hline
       \multirow{2}*{\textbf{Metric}} &\multirow{2}*{\textbf{Language}} &\multicolumn{3}{c|}{\textbf{Mean}} &\multicolumn{3}{c|}{\textbf{Median}} &\multicolumn{3}{c|}{\textbf{Comparison}}\\
        \cline{3-11}

 & &Statement &Branch &Method &Statement &Branch &Method &Statement \textit{vs} Branch &Statement \textit{vs} Method &Branch \textit{vs} Method \\
            \hline
 \multirow{4}*{APFD}
 &C	                 
 & 0.89 & 0.90 & 0.87 & 0.88 & 0.89 & 0.87	
 &1.65E-15/0.48	& 0/0.62        &0/0.65	 \\\cline{2-11}
 
&Java (test-method)	
& 0.68 & 0.69 & 0.68 & 0.69 & 0.73 & 0.69	
&7.13E-97/0.43	&2.04E-3/0.51	&1.39E-137/0.58	\\\cline{2-11}

&Java (test-class)	
& 0.54 & 0.55 & 0.54 & 0.54 & 0.54 & 0.54	
&4.34E-6/0.43	&4.34E-06/0.49	&1.67E-94/0.56	\\\cline{2-11}

&All	
& 0.72 & 0.73 & 0.71 & 0.72 & 0.75 & 0.70 	
&1.8E-24/0.48	&6.71E-34/0.52	&2.2E-113/0.54	\\\hline

  \end{tabular}}
\end{table*}

\subsection{RQ2: Impact of Code Coverage Granularity}
In our study, three basic structural coverage criteria (i.e., statement, branch and method) are adopted to evaluate the performance of proposed TCP techniques.
Previous empirical studies have demonstrated that 
different code coverage granularities may affect the APFD results \cite{2016Luo, 2020Huang}.
Thus, in this section, we examine how the selection of code coverage granularity influences the effectiveness of {\toolname}.

Figure \ref{FIG:SBM} presents the APFD results of {\toolname} for the three types of code coverage, according to the subject programs' language or test case granularity.
The language or test case granularity is shown on the $x$-axis and the APFD scores on the left $y$-axis. 
It can be observed that for C programs, statement and branch coverage are very considerable, and are more effective than method coverage. 
However, for Java programs, they have similar performance.

Table \ref{TAB:SBM} presents a comparison of the mean and median APFD values, and also shows the $p$-values/effect size $\hat{\textrm{A}}_{12}$ for the different code coverage granularity comparisons.
Column "C", "Java (test-method)", "Java (test -class)" and "All" is calculated based on all APFD values for C programs, Java programs at the test-method granularity, Java programs at the test-class granularity and all programs.
It can be observed that the APFD values are similar,
with the maximum mean and median value differences being less than 3\%, and less than 8\%, respectively.
According to the statistical comparisons, there is no single best code coverage type for {\toolname}, with each type sometimes achieving the best results.
Nevertheless, branch coverage appears slightly more effective than statement and method coverage for {\toolname}.


\finding{2}{
Overall, our analysis on the code coverage granularity reveals that
the code coverage granularity may only provide a small impact on {\toolname} testing effectiveness,
with branch coverage possibly slightly outperforming statement and method coverage.
}

\subsection{RQ3: Impact of Test Case Granularity}
In our study, the Java programs have two granularities of test cases (i.e., the test-class and test-method).
Following to previous studies \cite{2013Zhang, 2020Huang}, we also consider the test case granularity as a factor in the evaluation.
Thus, in this section, we also investigate how the test case granularity influence the effectiveness of {\toolname}.

The comparisons are presented in Figure \ref{FIG:SBM}.
{\toolname} usually has significantly lower average APFD values for prioritizing test cases at the test-class granularity than at the test-method granularity.

Table \ref{TAB:SBM} presents the statistical effectiveness comparisons of APFD between different granularities for {\toolname}.
Each cell in the Mean, Median, and Comparison columns represents the mean APFD value, the median value, and the $p$-values/effect size $\hat{\textrm{A}}_{12}$ for the different code coverage granularity comparisons, respectively.
Considering all the Java programs, as can be seen in Table \ref{TAB:SBM},
the mean and median APFD values at the test-method granularity are much higher than at the test-class granularity with all code coverage granularities.
In fact, the test case at the test-class granularity consists of a number of test cases at the test-method granularity.
For example, there exist 1000 test cases at the test-class granularity and more than 5000 test cases at test-method granularity for Java programs at Table \ref{TAB:programs}, resulting in a much larger number of test cases at the test-method granularity.
Thus, the permutation space of candidate test cases at the test-method granularity may be greater, which leads to a better fault detection rate \cite{2013Zhang}.


\finding{3}{
Overall, our analysis on the test case granularity reveals that
{\toolname} has better effectiveness performance when prioritizing test cases at the test-method granularity than at the test-class granularity in terms of fault detection rate.
}

\subsection{RQ4: Efficiency of {\toolname}}

\renewcommand\arraystretch{1.5}
\begin{sidewaystable*}[!tp]
\tiny
\centering
 \caption{\textbf{Efficiency: }Comparisons of execution costs in milliseconds for different TCP techniques.
 }
  \label{TAB:time}
  \setlength{\tabcolsep}{0.6mm}{
    \begin{tabular}{c|c|rrrrrrr|rrrrrrr|rrrrrrr}
     \hline
     
\multirow{2}*{\textbf{Language}} 
&\multirow{2}*{\textbf{Program}}  
&\multicolumn{7}{c|}{\textbf{Statement Coverage}} 
&\multicolumn{7}{c|}{\textbf{Branch Coverage}} 
&\multicolumn{7}{c}{\textbf{Method Coverage}} \\
        \cline{3-23}

         & &$\textit{TCP}_\textit{tot}$ 
         &$\textit{TCP}_\textit{add}$ 
         &$\textit{TCP}_\textit{lex}$ 
         &$\textit{TCP}_\textit{uni}$
         &$\textit{TCP}_\textit{art}$ &$\textit{TCP}_\textit{search}$ &$\textit{TCP}_\textit{\MakeLowercase{\toolname}}$  
        &$\textit{TCP}_\textit{tot}$ 
         &$\textit{TCP}_\textit{add}$ 
         &$\textit{TCP}_\textit{lex}$ 
         &$\textit{TCP}_\textit{uni}$
         &$\textit{TCP}_\textit{art}$ &$\textit{TCP}_\textit{search}$ &$\textit{TCP}_\textit{\MakeLowercase{\toolname}}$
         &$\textit{TCP}_\textit{tot}$ 
         &$\textit{TCP}_\textit{add}$ 
         &$\textit{TCP}_\textit{lex}$ 
         &$\textit{TCP}_\textit{uni}$
         &$\textit{TCP}_\textit{art}$ &$\textit{TCP}_\textit{search}$ &$\textit{TCP}_\textit{\MakeLowercase{\toolname}}$
         \\
            \hline

\multirow{5}*{C} & $gnu\_flex$ & \textbf{4.31} & 368.5 & 1028.03 & 687.98 & 7937.7 & 4189.23 & \textbf{138.69} 
& \textbf{2.66} & 274.83 & 565.48 & 365.06 & 7476.21 & 3846.69 & \textbf{79.3} 
& \textbf{0.60} & 60.31 & 45.87 & 31.26 & 353.5 & 3001.22 & \textbf{12.78} \\ 
 & $gnu\_grep$ & \textbf{2.5} & 164.06 & 469.2 & 312.44 & 3879.79 & 2834.07 & \textbf{48.18 }
 & \textbf{2.02} & 161.33 & 399.76 & 236.09 & 5442.39 & 2979.37 & \textbf{41.39} 
 &\textbf{ 0.45} & 38.94 & 27.27 & 17.54 & 213.55 & 2424.7 & \textbf{12.73} \\ 
 & $gnu\_gzip$ & \textbf{0.49} & \textbf{8.29} & 19.68 & 22.1 & 63.49 & 438.9 & 8.34 
 & \textbf{0.26} & 5.71 & 14.38 & 13.76 & 46.42 & 406.68 & \textbf{4.93} 
 & \textbf{0.06} & 2.71 & 1.81 & 1.85 & 6.72 & 349.84 & \textbf{2.00} \\ 
 & $gnu\_make$ & \textbf{0.82} & \textbf{13.07} & 40.78 & 40.36 & 101.77 & 556.23 & 13.25
 & \textbf{0.75} & 8.91 & 27.91 & 25.95 & 106.16 & 442.69 & \textbf{6.9} 
 & \textbf{0.06} & \textbf{1.25} & 2.34 & 2.25 & 5.7 & 205.53 & 1.66 \\ 
 & $gnu\_sed$ & \textbf{0.75} & 46.54 & 117.93 & 76.77 & 975.4 & 1952.64 & \textbf{12.78} 
 & \textbf{0.49} & 37.0 & 71.59 & 38.71 & 859.63 & 1949.83 & \textbf{7.9} 
 & \textbf{0.16} & 15.19 & 12.79 & 7.25 & 84.42 & 1766.16 & \textbf{4.01} \\ 
\hline 
\multirow{19}*{\shortstack{Java \\ (test-method)}} & $ant\_v1$ & \textbf{1.75} & 40.04 & 158.52 & 57.01 & 6185.81 & 29490.59 & \textbf{6.85} 
& \textbf{0.49} & 9.17 & 42.45 & 12.09 & 565.6 & 8852.76 & \textbf{2.36} 
& \textbf{0.29} & 7.12 & 28.49 & 8.69 & 1124.89 & 9018.78 & \textbf{2.22} \\ 
 & $ant\_v2$ & \textbf{4.9} & 179.95 & 618.03 & 220.52 & 37120.98 & 35375.86 & \textbf{22.17} 
 & \textbf{1.34} & 44.05 & 164.34 & 51.6 & 1645.86 & 26382.14 & \textbf{6.77} 
 & \textbf{0.78} & 32.07 & 103.95 & 32.92 & 6315.33 & 21068.79 & \textbf{8.24} \\ 
 & $ant\_v3$ & \textbf{4.98} & 181.64 & 616.08 & 217.66 & 36575.56 & 35217.32 & \textbf{22.03 }
 & \textbf{1.21} & 45.43 & 164.08 & 50.47 & 1618.44 & 20623.5 & \textbf{6.85} 
 & \textbf{0.84} & 32.36 & 104.01 & 32.83 & 6281.18 & 15832.8 & \textbf{8.41} \\ 
 & $ant\_v4$ & \textbf{30.26} & 1937.49 & 6928.77 & 2470.94 & 531511.88 & 403666.85 & \textbf{177.07 }
 & \textbf{7.79} & 533.67 & 1946.62 & 666.16 & 19113.81 & 104579.16 & \textbf{61.56} 
 & \textbf{4.41} & 352.97 & 1118.36 & 411.38 & 111594.23 & 65434.09 & \textbf{101.87} \\ 
 & $ant\_v5$ & \textbf{33.51} & 2339.8 & 8259.94 & 2980.85 & 701415.88 & 462601.72 & \textbf{210.64} 
 & \textbf{8.58} & 648.16 & 2316.6 & 795.81 & 22050.04 & 115193.82 & \textbf{71.31}
 & \textbf{5.11} & 424.71 & 1329.35 & 500.8 & 136199.85 & 75283.14 & \textbf{128.75} \\ 
 & $ant\_v6$ & \textbf{32.98} & 2261.22 & 8056.38 & 2899.7 & 675083.49 & 462799.01 & \textbf{204.27} \textbf{}& \textbf{9.04} 
 & 635.54 & 2271.12 & 779.82 & 18858.47 & 102967.53 & \textbf{70.19} 
 & \textbf{4.98} & 417.81 & 1294.7 & 488.7 & 127806.2 & 73827.57 & \textbf{125.04} \\ 
 & $ant\_v7$ &\textbf{ 75.39} & 6447.83 & 28863.39 & 10719.12 & 3582724.65 & 1243589.12 & \textbf{608.8} 
 & \textbf{21.54} & 2449.7 & 8276.29 & 3030.26 & 103270.76 & 302281.19 & \textbf{222.36} 
 & \textbf{12.84} & 1577.69 & 4793.59 & 1913.45 & 697296.49 & 135253.39 & \textbf{486.84} \\ 
 & $ant\_v8$ & \textbf{71.15} & 6576.87 & 29071.21 & 10645.15 & 3588748.59 & 1225224.6 & \textbf{637.39} 
 & \textbf{20.11} & 1985.4 & 8372.64 & 3052.9 & 104345.4 & 256010.28 & \textbf{225.68} 
 & \textbf{12.13} & 1637.23 & 4832.87 & 1911.08 & 701421.34 & 175257.75 & \textbf{579.66} \\ 
 & $jtopas\_v1$ & \textbf{2.05} & 37.08 & 157.98 & 52.08 & 5590.25 & 49553.33 & \textbf{10.79} 
 & \textbf{0.66} & 13.76 & 56.64 & 18.1 & 2054.62 & 22940.51 & \textbf{3.86} 
 & \textbf{0.14 }& 3.34 & 12.28 & 3.4 & 528.61 & 3118.26 & \textbf{1.98} \\ 
 & $jtopas\_v2$ & \textbf{1.97} & 39.16 & 164.99 & 55.08 & 6068.39 & 59249.46 & \textbf{11.3} 
 & \textbf{0.67} & 15.49 & 59.5 & 19.12 & 2170.79 & 47675.36 & \textbf{3.91} 
 & \textbf{0.16} & 3.74 & 13.44 & 3.9 & 576.68 & 8540.78 & \textbf{2.54} \\ 
 & $jtopas\_v3$ & \textbf{6.98} & 166.64 & 714.39 & 228.32 & 57582.22 & 65410.86 & \textbf{36.91} 
 & \textbf{2.34} & 59.67 & 257.85 & 78.34 & 14679.05 & 60761.66 & \textbf{14.23} 
 & \textbf{0.43} &\textbf{} 16.93 & 54.2 & 16.77 & 3955.14 & 32994.5 & \textbf{6.76} \\ 
 & $jmeter\_v1$ & \textbf{0.44} & 4.94 & 27.79 & 6.18 & 346.59 & 31128.38 & \textbf{1.58} 
 & \textbf{0.18} & 1.02 & 6.16 & 1.23 & 25.68 & 5259.7 & \textbf{0.44} 
 & \textbf{0.14} & 0.89 & 4.88 & 1.01 & 68.12 & 5013.62 & \textbf{0.44} \\ 
 & $jmeter\_v2$ & 1.12 & 10.99 & 68.28 & 13.83 & 1063.24 & 25942.38 & \textbf{3.12} 
 & \textbf{0.24} & 1.73 & 9.64 & 2.07 & 47.21 & 6825.12 & \textbf{0.62} 
 & \textbf{0.25} & 2.09 & 10.6 & 2.2 & 185.66 & 7374.39 & \textbf{0.86} \\ 
 & $jmeter\_v3$ & \textbf{1.39} & 14.1 & 81.6 & 17.61 & 1380.78 & 17542.05 & \textbf{3.78} 
 & \textbf{0.16} & 1.93 & 10.85 & 2.39 & 58.86 & 4833.83 & \textbf{0.7} 
 & \textbf{0.19} & 2.94 & 15.61 & 3.51 & 295.78 & 2938.2 &\textbf{ 0.97} \\ 
 & $jmeter\_v4$ & \textbf{1.12} & 16.06 & 83.42 & 18.04 & 1422.88 & 11761.66 & \textbf{3.81} 
 & \textbf{0.16} & 2.1 & 10.86 & 2.4 & 59.07 & 1481.61 & \textbf{0.67} 
 &\textbf{ 0.21} & 3.3 & 15.66 & 3.65 & 303.56 & 1934.82 & \textbf{1.01} \\ 
 & $jmeter\_v5$ & \textbf{1.24} & 18.35 & 96.51 & 21.21 & 1801.14 & 13119.46 & \textbf{4.42} 
 & \textbf{0.23} & 2.47 & 12.67 & 2.84 & 71.5 & 1347.98 & \textbf{0.77} 
 & \textbf{0.42} & 3.69 & 18.32 & 4.47 & 377.77 & 2001.66 & \textbf{1.20} \\ 
 & $xmlsec\_v1$ & \textbf{1.13} & 22.06 & 99.16 & 22.64 & 2365.41 & 10392.07 & \textbf{4.20} 
 & \textbf{0.47} & 4.69 & 24.3 & 5.52 & 118.31 & 1368.55 & \textbf{1.20} 
 & \textbf{0.25} & 3.35 & 14.51 & 3.06 & 339.86 & 1435.36 & \textbf{1.16} \\ 
 & $xmlsec\_v2$ & \textbf{1.21} & 24.06 & 108.63 & 26.1 & 2703.6 & 11188.71 & \textbf{4.97} 
 & \textbf{0.54} & 6.37 & 28.58 & 7.03 & 150.92 & 1593.97 & \textbf{1.54} 
 & \textbf{0.18} & 3.54 & 15.26 & 3.34 & 375.93 & 1363.08 & \textbf{1.30} \\ 
 & $xmlsec\_v3$ & \textbf{1.21} & 18.78 & 89.32 & 20.26 & 2098.82 & 10128.92 & \textbf{4.45} 
 & \textbf{0.36} & 5.12 & 24.59 & 5.69 & 111.92 & 1314.03 & \textbf{1.28 }
 & \textbf{0.14} & 2.7 & 12.03 & 2.45 & 259.3 & 1050.24 & \textbf{1.03} \\ 
\hline 
\multirow{19}*{\shortstack{Java \\ (test-class)}} & $ant\_v1$ & \textbf{0.42} & 3.69 & 19.67 & 4.33 & 133.56 & 6457.16 & \textbf{1.44} 
& \textbf{0.13} & 0.82 & 5.71 & 1.01 & 35.04 & 1178.75 & \textbf{0.40} 
& \textbf{0.10} & 0.62 & 3.86 & 0.73 & 25.2 & 922.62 & \textbf{0.33} \\ 
 & $ant\_v2$ & \textbf{1.36} & 11.57 & 64.61 & 14.26 & 628.79 & 18689.06 & \textbf{3.65} 
 & \textbf{0.44} & 2.93 & 17.66 & 3.64 & 171.69 & 3686.37 & \textbf{1.01} 
 & \textbf{0.19} & 1.78 & 11.26 & 2.43 & 111.20 & 2376.55 & \textbf{0.72} \\ 
 & $ant\_v3$ & \textbf{1.4} & 11.37 & 64.0 & 14.28 & 630.27 & 19582.33 & 3.85 & 
 \textbf{0.27} & 2.87 & 17.52 & 3.58 & 170.01 & 4371.02 & \textbf{1.06} 
 & \textbf{0.28} & 1.79 & 10.98 & 2.29 & 109.12 & 2755.49 & \textbf{0.70} \\ 
 & $ant\_v4$ & \textbf{5.76} & 76.96 & 420.2 & 101.54 & 9003.24 & 102183.09 & \textbf{17.85} 
 & \textbf{1.57} & 22.47 & 119.32 & 26.76 & 1933.59 & 19413.74 & \textbf{4.94} 
 & \textbf{1.22} & 13.56 & 64.98 & 15.84 & 1465.66 & 13110.27 &\textbf{ 2.90} \\ 
 & $ant\_v5$ & \textbf{6.44} & 88.95 & 460.8 & 109.61 & 10078.73 & 116900.69 & \textbf{19.14} 
 & \textbf{1.72} & 24.71 & 131.13 & 28.89 & 2139.8 & 20867.59 & \textbf{5.34} 
 & \textbf{0.9} & 16.01 & 70.96 & 16.95 & 1656.95 & 13822.1 & \textbf{3.12} \\ 
 & $ant\_v6$ & \textbf{6.65} & 82.74 & 451.09 & 107.43 & 9705.8 & 116659.29 & \textbf{18.64} 
 & \textbf{1.84} & 22.4 & 128.29 & 27.67 & 2017.54 & 21966.04 & \textbf{5.14} 
 & \textbf{0.89} & 13.27 & 69.35 & 16.64 & 1587.45 & 15390.26 & \textbf{3.02 }\\ 
 & $ant\_v7$ & \textbf{12.8} & 249.01 & 1193.98 & 326.64 & 131530.77 & 163839.7 & \textbf{39.39} 
 & \textbf{3.60} & 69.75 & 344.87 & 90.07 & 5848.22 & 54516.25 & \textbf{11.67} 
 & \textbf{2.00} & 42.48 & 196.28 & 53.68 & 7277.59 & 32652.62 & \textbf{7.48} \\ 
 & $ant\_v8$ & \textbf{12.68} & 247.24 & 1196.01 & 325.44 & 132814.84 & 156276.32 & \textbf{39.79} 
 & \textbf{3.61} & 70.38 & 346.6 & 90.6 & 5934.71 & 57982.7 & \textbf{11.84 }
 & \textbf{1.97} & 41.67 & 195.86 & 53.18 & 7261.99 & 41400.42 & \textbf{7.17} \\ 
 & $jtopas\_v1$ & \textbf{0.15} & 0.62 & 4.66 & 0.74 & 6.16 & 8268.77 & \textbf{0.50} 
 & \textbf{0.06} & 0.20 & 1.73 & 0.28 & 2.29 & 2093.0 & \textbf{0.19} 
 & \textbf{0.02} & 0.05 & 0.42 & 0.05 & 0.53 & 677.81 & \textbf{0.04} \\ 
 & $jtopas\_v2$ & \textbf{0.16} & 0.61 & 5.34 & 0.86 & 7.67 & 7132.1 & \textbf{0.55} 
 & \textbf{0.06} & 0.22 & 1.98 & 0.37 & 2.95 & 2625.9 & \textbf{0.20} 
 & \textbf{0.02} & \textbf{0.05} & 0.48 & 0.08 & 0.78 & 898.9 &\textbf{ 0.05} \\ 
 & $jtopas\_v3$ & \textbf{0.49} & 2.12 & 15.62 & 2.65 & 31.87 & 27084.71 & \textbf{1.38} 
 & \textbf{0.19} & 0.72 & 5.63 & 0.88 & 11.73 & 6509.74 & \textbf{0.48} 
 & \textbf{0.07} & 0.16 & 1.36 & 0.22 & 3.02 & 1514.9 & \textbf{0.13} \\ 
 & $jmeter\_v1$ & \textbf{0.2} & 0.77 & 5.99 & 0.89 & 16.66 & 6560.88 & \textbf{0.44} 
 & \textbf{0.06} & 0.15 & 1.33 & 0.19 & 3.66 & 1602.51 &\textbf{ 0.10} 
 & \textbf{0.04} & 0.16 & 1.19 & 0.18 & 3.36 & 1254.53 & \textbf{0.09} \\ 
 & $jmeter\_v2$ & \textbf{0.31} & 1.32 & 12.27 & 1.81 & 38.84 & 10595.78 & \textbf{0.79} 
 & \textbf{0.07} & 0.24 & 1.85 & 0.28 & 6.11 & 1655.58 & \textbf{0.14} 
 &\textbf{ 0.05} & 0.26 & 2.06 & 0.34 & 6.66 & 1272.91 & \textbf{0.18 }\\ 
 & $jmeter\_v3$ & \textbf{0.36} & 1.97 & 16.52 & 2.57 & 64.97 & \textbf{}13330.05 
 & \textbf{1.02} & \textbf{0.07} & 0.3 & 2.57 & 0.44 & 9.54 & 1464.98 & \textbf{0.16} 
 & \textbf{0.08} & 0.44 & 3.37 & 0.53 & 14.07 & 1915.9 & \textbf{0.25} \\ \textbf{}
 & $jmeter\_v4$ & \textbf{0.36} & 1.98 & 16.94 & 2.64 & 67.29 & 13366.85 & \textbf{1.05} 
 & \textbf{0.06} & 0.28 & 2.59 & 0.31 & 9.44 & 2238.46 & \textbf{0.17} 
 & \textbf{0.09} & 0.42 & 3.44 & 0.54 & 14.48 & 2167.08 & \textbf{0.25} \\ 
 & $jmeter\_v5$ & \textbf{0.43} & 2.43 & 20.95 & 3.28 & 93.95 & 10861.39 & \textbf{1.30 }
 & \textbf{0.07} & 0.36 & 3.02 & 0.51 & 11.7 & 2088.49 & \textbf{0.20} \textbf{}
 & \textbf{0.08} & 0.53 & 4.19 & 0.66 & 19.83 & 2475.02 & \textbf{0.29} \\ 
 & $xmlsec\_v1$ & \textbf{0.21} & 1.02 & 8.25 & 1.21 & 18.86 & 6315.35 & \textbf{0.65 }
 & \textbf{0.06} & 0.24 & 2.14 & 0.3 & 4.15 & 576.82 & \textbf{0.18} 
 & \textbf{0.04} & 0.16 & 1.3 & 0.16 & 2.79 & 875.06 & \textbf{0.10} \\ 
 & $xmlsec\_v2$ & \textbf{0.2 }& 1.01 & 8.53 & 1.24 & 19.05 & 6104.03 & \textbf{0.64 }
 &\textbf{ 0.09} & 0.25 & 2.47 & 0.32 & 4.31 & 1269.4 & \textbf{0.18} 
 & \textbf{0.06} & 0.14 & 1.22 & 0.18 & 2.72 & 937.54 & \textbf{0.11} \\ 
 & $xmlsec\_v3$ & \textbf{0.17} & 0.77 & 6.9 & 0.98 & 12.84 & 4190.4 & \textbf{0.57} 
 & \textbf{0.06} & 0.21 & 2.01 & 0.26 & 2.93 & 521.64 & \textbf{0.14} 
 & \textbf{0.04} & 0.11 & 1.01 & 0.14 & 1.8 & 706.94 & \textbf{0.08} \\ 
\hline 
\multicolumn{2}{c|}{\textbf{\textit{Average}} } & \textbf{7.77} & 505.2 & 2091.45 & 764.06 & 222084.94 & 116924.45 & \textbf{54.71} 
& \textbf{2.24} & 166.81 & 611.01 & 221.85 & 7517.83 & 30431.32 & \textbf{20.47} 
&\textbf{ 1.24} & 111.17 & 337.81 & 129.6 & 42221.74 & 18223.62 & \textbf{35.36} \\ 

\hline


  \end{tabular}}
\end{sidewaystable*}
In this section, to evaluate the efficiency of {\toolname}, we calculate the execution time for all TCP techniques.
Table \ref{TAB:time} presents the statistics about time costs (i.e., the preprocessing time and prioritization time) for all subject programs and studied TCP techniques.

Specifically, the preprocessing time contains the compilation time for executing the program and the instrumentation time for collecting the coverage information.
Thus the preprocessing time of the subject programs is the same for different TCP techniques and is not presented.
Apart from the first two columns that display the program name and the programming language it belongs to, each cell in the table shows the mean prioritization time over the 1,000 independent runs using each TCP technique.

As discussed in Section \ref{sec:exp}, the Java programs have each version individually adapted to collect the code coverage information, with different versions using different test cases.
Thus, the prioritization time is collected for each Java program version. 
In contrast, each $P_{V0}$ version of the C programs is compiled and instrumented to collect the code coverage information for each test case, and all program versions use the same test cases.
Thus, each C program version has the same prioritization time.
As a result, we present the time costs for each Java program version and C Program.
Furthermore, because all the studied TCP techniques prioritize test cases after the coverage information is collected,
they are all deemed to have the preprocessing time.

Based on the time costs, we have the following observations:
(1)
As expected, the time costs for all TCP techniques (including {\toolname}) are lowest with method coverage, followed by branch, and then statement coverage for all programs.
As shown in Table \ref{TAB:programs}, the number of methods is much smaller than the number of branches, which in turn is smaller than the number of statements.
Thus, the coverage information representing the related test cases is smaller, requiring less time to compute the priority value.
(2)
It is also expected that (for the Java programs) prioritization at the test-method granularity would take longer than at the test-class granularity, regardless of code coverage granularities.
As shown in Table \ref{TAB:programs}, the number of test cases to be prioritized at the test-method granularity is large than those at the test-class granularity, which requires more prioritization iterations.

(3)
{\toolname} requires much less time to prioritize test cases than most studied TCP techniques (e.g., $TCP_{add}$, $TCP_{lexi}$, $TCP_{uni}$, $TCP_{art}$ and $TCP_{search}$) irrespective of subject program, and code coverage and test case granularities.
Meanwhile, {\toolname} can improve the costs of $TCP_{add}$ by 85\% on average.
Considering that $TCP_{add}$ remains state-of-the-art, the decreases in costs can achieve slightly better or comparable fault detection effectiveness and thus are valuable actually.
It should be noted that $TCP_{tot}$ has a much faster prioritization rate than $TCP_{\MakeLowercase{\toolname}}$, as it does not use feedback information during the prioritization process. 
However, $TCP_{tot}$ performs worst among all TCP techniques and is usually considered as a low bound control TCP technique \cite{2013Zhang}.

\finding{4}{
Overall, our analysis on the efficiency reveals that
except $TCP_{tot}$, the APFD values of which is much lower than {\toolname}, 
$TCP_{\MakeLowercase{\toolname}}$ has much less time to prioritize test cases than $TCP_{add}$, $TCP_{lexi}$, $TCP_{uni}$, $TCP_{art}$ and $TCP_{search}$.
}


\section{Related Work}
\label{sec:rw}
A considerable amount of research has been conducted to improve regression testing performance on various issues \cite{2019Lou,pan2022test,yu2021prioritize,sun2022path,mondal2021hansie,fang2014similarity,haghighatkhah2018test}.
We focus on the coverage-based TCP techniques and summarize the existing work from the following categories.

\subsection{Prioritization Strategies}
Despite the large body of research on coverage-based TCP \cite{fang2014similarity, 2018Miranda, 2015Epitropakis, 2020Peng}, the \textit{total-greedy} and \textit{additional-greedy} greedy strategies remain the most widely investigated prioritization strategies ~\cite{1999Rothermel}.
In addition to the above \textit{greedy-based} strategies, researchers have also investigated other generic strategies \cite{2007Li, 2009Jiang}.
For example, Li et al. \cite{2007Li} transform the TCP problem into a search problem and propose two \textit{search-based} prioritization strategies (i.e., a hill-climbing strategy and a genetic strategy).
Furthermore, inspired by the advantages of adaptive random testing (ART) in replacing random testing (RT) \cite{2019Huang}, Jiang et al. \cite{2009Jiang} investigate ART to improve random test case prioritization and propose an \textit{art-based} strategy based on the distribution of test cases across the input domain. 

Researchers have also proposed some alternative strategies in previous studies to take further advantage of the code coverage information (e.g., the covered times of code units).
For example, Eghbali et al. \cite{2016Eghbali} propose an enhanced \textit{additional-greedy} strategy for breaking ties using the notion of lexicographical ordering (i.e., the \textit{lexicographical-greedy} strategy), where fewer covered code units should have higher priority value for coverage. 
Specifically, unlike traditional \textit{greedy-based} strategies \cite{2000Elbaum}, Eghbali et al. do not categorize all code units into two distinct groups, (i.e., covered and not covered).
At each iteration, huge calculations are required to calculate the priority values of all code units based on the number of times they are covered, and then each one in the remaining test cases is lexicographically compared against others.
Similarly, Zhang et al. \cite{2013Zhang} propose a variant of the \textit{additional-greedy} strategy to unify the \textit{total-greedy} strategy and the \textit{additional-greedy} strategy.
Each time a code unit is covered by a test case, the probability that the code unit contains undetected faults is reduced by some ratio between 0\% (as in the \textit{total-greedy} strategy) and 100\% (as in the \textit{additional-greedy} strategy),
 which can also be considered as an effective strategy to break tie cases.
The above techniques attempt to make use of more accurate coverage information obtained by additional calculations.
For example, the \textit{lexicographical-greedy} strategy needs to rank all code units based on the number of times they are covered,
while the \textit{unified-greedy} strategy needs to calculate the value of fault detection probability for each code unit.
Most recently, Zhou et al. \cite{2021Zhou} propose eight parallel test prioritization techniques, which are adapted from four typical sequential test prioritization techniques (including \textit{total-greedy}, the \textit{additional-greedy} strategy , the \textit{search-based} strategy, and the \textit{art-based} strategy).
Different from traditional sequential TCP techniques, it aims at generating a set of test sequences, each of which is allocated in an individual computing resource.
Thus, we do not include it in this work.


It can be observed that most existing TCP strategies tend to consider the whole candidate set in prioritization iterations.
To address this limitation, we pay attention to partial candidate test cases with the aid of previous priority values, resulting in a better performance in both effectiveness and efficiency.


\subsection{Coverage criteria}
In principle, TCP techniques can use any test adequacy criterion as the underlying coverage criterion \cite{2016Hao, fang2012comparing}.
Among various criteria, structural coverage has been widely adopted in in previous TCP research, such as statement coverage\cite{1999Rothermel, 2020Nucci}, branch coverage\cite{2009Jiang}, method coverage \cite{2013Zhang,2017Wang}, block coverage\cite{2007Li} and modified condition/decision coverage~\cite{2003Jones}.
Elbaum et al.~\cite{2000Elbaum} also propose a fault-exposing-potential (FEP) criterion based on the probability of the test case detecting a fault. 
Fang et al.~\cite{fang2012comparing} use logic coverage for TCP, where high coverage of logic expressions implies a high probability of detecting faults.
Recently, Chi et al.~\cite{2018Chi} demonstrate that basic structural coverage may not be enough to predict fault detection capability and propose a dynamic relation-based coverage based on method call sequences.
Wang et al. \cite{2017Wang} detect fault-prone source code by existing code inspection techniques and then propose a quality-aware TCP technique (i.e., QTEP) by considering the weighted source code in terms of fault-proneness.
However, such techniques require not only coverage information but also other source code information (e.g., the defect prediction results and method call sequences) and thus are not considered in our work.
In this work, we investigate how the basic structure coverage criteria influence the performance of TCP techniques.

\subsection{Empirical studies}
As an effective regression testing technique, TCP has been extensively studied in the literature from both academic and industrial perspectives.
Recently, researchers also performed a large number of empirical studies to investigate TCP from different aspects.
For example, several studies usually focus on the performance of the traditional dynamic test prioritization regarding some effectiveness and efficiency criteria (e.g., $APFD$, $APFD_C$, and prioritization time) \cite{1999Rothermel,2000Elbaum,2006Do,2010Do}.
Meanwhile, Lu et al.~\cite{2016Lu} were the first to investigate how real-world software evolution impacts the performance of prioritization strategies:
They reported that source code changes have a low impact on the effectiveness of traditional dynamic techniques, but that the opposite was true when considering new tests in the process of evolution.
Citing a lack of comprehensive studies comparing static and dynamic test prioritization techniques, Luo et al.~\cite{2016Luo,2019Luo} compared static TCP techniques with dynamic ones. Henard et al.~\cite{2016Henard} compared white-box and back-box TCP techniques.
In this work, we focus on the coverage-based TCP techniques and conduct an extensive study to evaluate {\toolname} with six state-of-the-art TCP techniques.

\section{Threats to validity}
\label{sec:threats}
To facilitate the replication and verification of our experiments, we have made the relevant materials (including source code, subject programs, test suites, and mutants) available at \cite{myurl}.
Despite that, our study may still face some threats to validity.

\subsection{Internal validity}
The implementation of our experiment may introduce threats to internal validity.
First, randomness might affect the reliability of conclusions.
To address this, we repeat the prioritization process 1,000 times and used statistical analysis to assess the strategies.
Second, the data structures used in the prioritization algorithms, and the faults in the source code, may introduce noise when evaluating the effectiveness and efficiency.
To minimize these threats, we use data structures that are as similar as possible, and carefully reviewed all source code before conducting the experiment.
Third, to assess the effectiveness of TCP techniques, the most widely used metric $APFD$ is adopted in our experiment.
However, $APFD$ only reflects the rate at which faults are detected, ignoring the time and space costs.
Our future work will involve additional metrics (e.g., $APFD_C$) that can measure other practical performance aspects of prioritization strategies.

\subsection{External validity}
The main threat to external validity lies in the selection of the subject programs and faults.
First, although 19 Java and 30 C program versions with various sizes are adopted in our experiment, the results may not generalize to programs written in other languages (e.g., C++ and Python).
Meanwhile, the relative performances of TCP techniques on the used mutants may not be generalizable to the real faults, despite the fact that mutation testing have argued to be an appropriate approach for assessing TCP performance~\cite{2005Andrews,2005Do,2014Just}, 
To mitigate these threats, additional studies will be conducted to investigate the performance of TCP on programs with real faults and other languages in the future.

\section{Conclusion}
\label{sec:con}

In this paper, we have introduced a generic partial attention mechanism that adopts priority values of previously selected test cases to avoid considering all test cases.
We also apply the concept to the \textit{additional-greedy} strategy and implement a novel coverage-based TCP technique, \textit{partition ordering based prioritization} ({\toolname}).
Results from our empirical study have demonstrated that {\toolname} can achieve better fault detection rate than six state-of-the-arts (i.e., \textit{total-greedy}, \textit{additional-greedy}, \textit{unified-greedy}, \textit{lexicographical-greedy}, \textit{art-based}, and \textit{search-based} TCP techniques).
{\toolname} is also found to have much less prioritization time to prioritize test cases than most state-of-the-arts (except the \textit{total-greedy} strategy) and the improvement can achieve 85\% - 99\% on average.

In the future, we plan to continue refining the generic partial attention mechanism and extend it to other TCP techniques (e.g., the \textit{lexicographical-greedy} strategy).
We will also launch an extensive effort on understanding the impact of the proposed technique for other application domains of TCP research \cite{wang2021prioritizing,chen2020practical,sharif2021deeporder}, such as configuration testing\cite{2021Cheng, 2020Sun} and combinatorial testing \cite{2014Henard, 2020Wu}.

\section*{Acknowledgments}
The authors would like to thank the anonymous reviewers for insightful comments.
This research is partially supported by the National Natural Science Foundation of China (No. 61932012, 62141215) and Science, Technology and Innovation Commission of Shenzhen Municipality (CJGJZD20200617103001003).



\normalem
\bibliographystyle{elsarticle-num} 
\bibliography{reference}
\vspace{12pt}





\end{document}